\documentclass[UTF8,twocolumn]{aastex61}
\usepackage{amsmath}
\usepackage{CJK}

\newcommand{\ha}{\hbox{H$\alpha$}}
\newcommand{\hb}{\hbox{H$\beta$}}

\newcommand{\hd}{\hbox{H$\delta$}}

\newcommand{\gsim}{\lower.5ex\hbox{$\; \buildrel > \over \sim \;$}}
\newcommand{\lsim}{\lower.5ex\hbox{$\; \buildrel < \over \sim \;$}}

\newcommand{\oiii}{\hbox{[O\,{\sc iii}]}}
\newcommand{\nii}{\hbox{[N\,{\sc ii}]}}

\newcommand{\dn}{$D_n(4000)$}
\newcommand{\tauB}{$\tau_{\mathrm{B}}$}

\newcommand{\sigha}{$\mathrm{\Sigma_{H\alpha}}$}
\newcommand{\ewha}{$\mathrm{EW_{H\alpha}}$}

\submitjournal{AAS}

\shorttitle{Attenuation Curve for Local Star-forming Regions}
\shortauthors{Teklu et al.}

\begin{document}

\begin{CJK*}{UTF8}{gbsn}
\title{Dust Attenuation Curve for Local Subgalactic Star-forming Regions}

\email{berzaf12@mail.ustc.edu.cn, zesenlin@mail.ustc.edu.cn, xkong@ustc.edu.cn}
\author{Berzaf Berhane Teklu}
\affil{Key Laboratory for Research in Galaxies and Cosmology, Department of Astronomy, University of Science and Technology of China, Hefei 230026, China}
\affil{School of Astronomy and Space Sciences, University of Science and Technology of China, Hefei, 230026, China}

\author{Zesen Lin}
\altaffiliation{Berzaf Berhane Teklu and Zesen Lin contributed equally to this work.}
\affil{Key Laboratory for Research in Galaxies and Cosmology, Department of Astronomy, University of Science and Technology of China, Hefei 230026, China}
\affil{School of Astronomy and Space Sciences, University of Science and Technology of China, Hefei, 230026, China}

\author{Xu Kong}
\affil{Key Laboratory for Research in Galaxies and Cosmology, Department of Astronomy, University of Science and Technology of China, Hefei 230026, China}
\affil{School of Astronomy and Space Sciences, University of Science and Technology of China, Hefei, 230026, China}

\author{Enci Wang}
\affil{Department of Physics, ETH Zurich, Wolfgang-Pauli-strasse 27, CH-8093 Zurich, Switzerland}

\author{Yulong Gao}
\affil{Key Laboratory for Research in Galaxies and Cosmology, Department of Astronomy, University of Science and Technology of China, Hefei 230026, China}
\affil{School of Astronomy and Space Sciences, University of Science and Technology of China, Hefei, 230026, China}

\author{Qing Liu}
\affil{Key Laboratory for Research in Galaxies and Cosmology, Department of Astronomy, University of Science and Technology of China, Hefei 230026, China}
\affil{School of Astronomy and Space Sciences, University of Science and Technology of China, Hefei, 230026, China}

\author{Ning Hu}
\affil{Key Laboratory for Research in Galaxies and Cosmology, Department of Astronomy, University of Science and Technology of China, Hefei 230026, China}
\affil{School of Astronomy and Space Sciences, University of Science and Technology of China, Hefei, 230026, China}

\author{Haiyang Liu}
\affil{Key Laboratory for Research in Galaxies and Cosmology, Department of Astronomy, University of Science and Technology of China, Hefei 230026, China}
\affil{School of Astronomy and Space Sciences, University of Science and Technology of China, Hefei, 230026, China}

\begin{abstract}
  We compile a sample of about 157,000 spaxels from the Mapping Nearby Galaxies at the Apache Point Observatory (MaNGA) survey to derive the average dust attenuation curve for subgalactic star-forming regions of local star-forming galaxies (SFGs) in the optical wavelength, following the method of Calzetti et al. We obtain a \dn-independent average attenuation curve for spaxels with $1.1\leq D_n(4000)<1.3$, which is similar to the one derived from either local starbursts or normal SFGs. We examine whether and how the shape of the average attenuation curve changes with several local and global physical properties. For spaxels with $1.2\leq D_n(4000)<1.3$, we find no dependence on either local or global physical properties for the shape of the average attenuation curve. However, for spaxels with younger stellar population ($1.1\leq D_n(4000)<1.2$), shallower average attenuation curves are found for star-forming regions with smaller stellar mass surface density, smaller star formation rate surface density, or those residing in the outer region of galaxies. These results emphasize the risk of using one single attenuation curve to correct the dust reddening for all type of star-forming regions, especially for those with fairly young stellar population.
\end{abstract}


\section{Introduction}
\label{sec:intro}

  The spectral energy distributions (SEDs) of galaxies encode abundant information of their physical properties, such as star formation rate (SFR), star formation history (SFH), stellar mass ($M_*$), and so on (see \citealt{Conroy2013} for a review). However, as a main component of interstellar medium, dust can absorb or/and scatter starlight, resulting in a significant reddening of the intrinsic SEDs that prevents us from recovering the physical properties of galaxies. Such reddening, which is wavelength dependent and more serious at the blue end, can be described by an extinction (for point-like sources, e.g., \citealt{Cardelli1989,Misselt1999}) or attenuation (for extend sources, e.g., \citealt{Calzetti2000,Charlot2000}) curve \citep{Calzetti2001}.

  For the Milky Way and several local galaxies in which individual stars can be resolved, extinction curves were constructed to describe the dust effects including absorption and scattering out of the line of sight (e.g, \citealt{Cardelli1989,Misselt1999,Gordon2003,Clayton2015}). The behavior of extinction as a function of wavelength depends on the composition and the size distribution of dust grains, and can be used to retrieve the properties of the interstellar dust \citep{Weingartner2001,Draine2003}. For distant galaxies, attenuation curves are derived and take into account not only the extinction but also both the dust scattering into the line of sight and the geometry between stars and dust (e.g., \citealt{Calzetti2000,Battisti2016,Salmon2016,Narayanan2018}).

  Due to the crucial character of dust attenuation curve in investigation of galaxy formation and evolution, a number of efforts were undertaken to reveal the dust attenuation curve for either local (e.g., \citealt{Calzetti2000,Wild2011,Battisti2016,Battisti2017,Salim2018}) or high-redshift (e.g., \citealt{Scoville2015,Reddy2015,Salmon2016,LoFaro2017,Cullen2018,Tress2018}) galaxies. In the local universe, average attenuation curves were determined for both starburst galaxies \citep{Calzetti1994,Calzetti2000} and normal star-forming galaxies (SFGs; \citealt{Wild2011,Battisti2016,Battisti2017}). The shape of the attenuation curve was found to correlate with some physical properties of SFGs, such as inclination \citep{Wild2011,Battisti2017} and stellar mass surface densities \citep{Wild2011}. For individual galaxies, dust attenuation curves were also derived based on multiwavelength SED fitting from which a large diversity of both the slope and the amplitude of UV bump at 2175 \AA\ was revealed \citep{Salim2018}.

  At the subgalactic scale, as the development of the IFU observation, spatially resolved attenuation curves are available for early-type galaxies with dust lane. For instance, on the basis of the assumed simple dust distribution, \cite{Viaene2017} calculated spatially resolved optical dust attenuation curves down to a physical scale of 0.1 kpc for NGC 5626. A similar study was also performed for FCC 167 \citep{Viaene2019}.

  However, in most cases, due to the unknown and complex intrinsic SED of star-forming regions or SFGs, only the average attenuation curve can be derived. Given that most of attenuation curves reported by previous works were derived based on photometry/spectra of either the entire galaxies (e.g., \citealt{Calzetti1994,Wild2011,Salim2018}) or the most central regions of galaxies (e.g., \citealt{Battisti2016}), our knowledge to the wavelength dependence of dust attenuation at the scale of star-forming region is very limited. Therefore, it is important to determine an average attenuation curve for subgalactic regions and examine whether the curves vary with physical scale or local physical properties.

  In this work, we aim at using the spatially resolved spectra from the Mapping Nearby Galaxies at Apache Point Observatory (MaNGA; \citealt{Bundy2015}) survey to derive the dust attenuation curve at subgalactic scale. We follow the procedure applied in \cite{Calzetti1994} and \cite{Battisti2016} to measure the behavior of the optical dust attenuation curves, using a sample of 982 SFGs ($\sim 157,000$ spaxels), as a function of several local (e.g., surface density of stellar mass, surface density of SFR, gas-phase metallicity) and global (e.g., $M_*$, inclination) physical properties.

  This paper is organized as follows. Section \ref{sec:data} describes the data processing and sample selection of spaxels in SFGs. In Section \ref{sec:methods} we briefly review the method of deriving attenuation curve and apply it to the MaNGA data. Our main results are presented in Section \ref{sec:results}. We discuss effects of the assumed stars/dust geometry and diffuse ionized gas (DIG) in Section \ref{sec:discussion} and summarize in Section \ref{sec:summary}. Throughout this paper, we adopt a flat $\Lambda$CDM cosmology with $\Omega_\Lambda=0.7$, $\Omega_\mathrm{m}=0.3$, and $H_0=70$ km s$^{-1}$ Mpc$^{-1}$ and an initial mass function of \cite{Chabrier2003}.

\section{Data and Sample Selection}
\label{sec:data}

  \subsection{MaNGA Overview}
  \label{ssub:MaNGA}

    The MaNGA\footnote{\url{https://www.sdss.org/dr14/manga/}} survey aims to observe spatial resolved spectra for $\sim 10,000$ galaxies across a redshift range of $0.01<z<0.15$. Spectra observed by the MaNGA survey have a wavelength range from 3600 to 10300 \AA\ with a spectral resolution from $R\sim1400$ at 4000 \AA\ to $R\sim2600$ near 9000 \AA. The FWHM of the reconstructed point spread function (PSF) of the data cubes is about 2\farcs5 \citep{Law2016}. In this work, we take the MaNGA data from the SDSS DR14\footnote{\url{https://www.sdss.org/dr14/}} \citep{Abolfathi2018} as the parent sample, which provides 2812 data cubes.

  \subsection{Data Processing}
  \label{ssub:data_processing}

    We first correct the Galactic foreground extinction based on the color excesses saved in the header of data cubes using the Milky Way dust extinction curve of \cite{Cardelli1989}. The underlying continua of spectra are modeled via the public full-spectrum fitting code of STARLIGHT \citep{CidFernandes2005}. During the fitting, a dust attenuation curve of \cite{Calzetti2000} and spectrum templates from the \cite{Bruzual2003} models are adopted. The best-fit continua are then subtracted from the original spectra to obtain the pure emission line spectra. We stress that the best-fit continua are only used to create the pure emission line spectra and thus measure emission line fluxes, while the derivation of attenuation curves described in Section \ref{ssec:application_to_IFU} is based on the observed spectra from which emission lines have been subtracted. Therefore, the derived emission line fluxes, as well as the derived attenuation curves, are not affected by using a \cite{Calzetti2000} attenuation curve for the continuum fitting.

    Each emission line is fitted with single Gaussian profile utilizing the MPFIT IDL code \citep{Markwardt2009}. The signal-to-noise ratios (S/N) of emission lines are estimated following the procedure of \cite{Ly2014}. We use the Balmer decrement to correct the galactic internal reddening, assuming an intrinsic flux ratio of $\ha/\hb=2.86$ under the Case B recombination with electron temperature of $T_{\mathrm{e}}=10,000$ K and electron density of $n_{\mathrm{e}}=100$ cm$^{-3}$ \citep{Storey1995}, together with the \cite{Calzetti2000} attenuation curve.

    The global stellar mass ($M_{*}$) and axial ratio ($b/a$) used in the following analysis are taken from the NASA-Sloan Atlas (NSA) catalog adopted by the MaNGA survey\footnote{\url{https://www.sdss.org/dr14/manga/manga-target-selection/nsa/}} \citep{Blanton2011}.

  \subsection{Sample Selection}
  \label{ssub:selection}

    From the 2812 galaxies released by the SDSS DR14, we first select SFGs based on the criterion of $\mathrm{NUV}-r<4$ that is widely used to select galaxies within the blue cloud in the literature (e.g., \citealt{Li2015}). To ensure reliable measurement of emission line fluxes, only spaxels with S/N(\ha) $>$ 5, S/N(\hb) $>$ 5, S/N(\oiii$\lambda5007$) $>$ 5, S/N(\nii$\lambda6583$) $>$ 3, and the S/N of continua greater than 10 are used. The excitation mechanisms of spaxels can be classified by the Baldwin--Phillips--Terlevich diagram (\citealt{Baldwin1981}), together with theoretical \citep{Kewley2001} or empirical \citep{Kauffmann2003a,Kewley2006} boundaries. Due to the fact that we only focus on star-forming regions, spaxels affected by the active galactic nucleus (AGN) are excluded from our sample by applying the \cite{Kauffmann2003a} demarcation. The above selections result in a sample of $\sim 326,000$ spaxels from 1227 SFGs.

\section{METHODS}
\label{sec:methods}

  \subsection{Basic Idea}
  \label{ssec:basis}

    As elucidated in \cite{Kinney1994}, for galaxies with similar morphology, similar SFHs but different inclinations, and thus different dust attenuation, the reddening curve can be derived from the comparison of their spectra. This method was generalized by \cite{Calzetti1994}, and is widely used in calculation of dust attenuation curve for local (e.g., \citealt{Battisti2016,Battisti2017a}) and high-redshift (e.g., \citealt{Reddy2015}) galaxies. In this work, we apply this method to our selected sample to obtain dust attenuation curve for subgalactic star-forming regions.

    Here, we give a brief description of the method, more discussions can be found in \cite{Calzetti1994} and \cite{Battisti2016}. This method begins with three main arguments/assumptions as following:
    \begin{itemize}
        \item The dust attenuation is dominated by a uniform foreground-like dust component.
        \item The reddening of ionized gas is correlated with that of the stellar continuum.
        \item Galaxies/regions have roughly similar SFHs and stellar populations.
    \end{itemize}

    The first assumption gives the relation between the observed and intrinsic fluxes as
    \begin{equation}\label{eq:fobs}
    f_{\lambda,\mathrm{obs}} = f_{\lambda,\mathrm{int}}e^{-\tau_{\lambda}},
    \end{equation}
    in which $f_{\lambda,\mathrm{obs}}$ and $f_{\lambda,\mathrm{obs}}$ are the observed and intrinsic fluxes at wavelength $\lambda$, respectively, $\tau_{\lambda}$ is the optical depth along the line of sight. Based on this formula, one can define the Balmer optical depth (\tauB) as the difference between optical depth at wavelength of \ha\ and \hb, i.e,
    \begin{equation}
    \tau_{\mathrm{B}} = \tau_{\mathrm{H\alpha}}-\tau_{\mathrm{H\beta}} = \ln\left(\frac{f_{\mathrm{H\alpha, obs}}/f_{\mathrm{H\beta,obs}}}{2.86}\right),
    \end{equation}
    where $\tau_{\mathrm{H\alpha}}$ ($\tau_{\mathrm{H\beta}}$) and $f_{\mathrm{H\alpha, obs}}$ ($f_{\mathrm{H\beta, obs}}$) are the optical depth and observed flux at wavelength of \ha\ (\hb), respectively. The ratio between the intrinsic fluxes of \ha\ and \hb\ is already replaced by 2.86 that is the intrinsic \ha-to-\hb\ flux ratio under the Case B recombination \citep{Storey1995}. If a specified total-to-selective extinction curve $k_{\lambda}\equiv A_{\lambda}/E(B-V)$ is applied, the color excess related to the ionized gas can be expressed as
    \begin{equation}
      E(B-V)_{\mathrm{gas}}=\frac{1.086\tau_{\mathrm{B}}}{k_{\mathrm{H\beta}}-k_{\mathrm{H\alpha}}},
    \end{equation}
    in which $k_{\mathrm{H\alpha}}$ ($k_{\mathrm{H\beta}}$) is the $k_{\lambda}$ value at the wavelengths of \ha\ (\hb). Hence, the Balmer optical depth can directly link to the nebular color excess $E(B-V)_{\mathrm{gas}}$. The second argument allows us to use nebular attenuation derived from the Balmer decrement to trace the stellar attenuation. Previous works showed that the color excesses derived from emission lines and stellar continuum  are correlated with each other in either galactic (e.g., \citealt{Calzetti1997}) or subgalactic (e.g., \citealt{Kreckel2013,Lin2020}) star-forming systems. Therefore, one can use \tauB\ to represent the relative reddening of of stellar continuum.

    Considering two star-forming regions with the same SFHs, and thus very similar intrinsic spectra, the only difference in their observed spectra should be due to dust attenuation. Denoting these two spectra as $f_{\lambda,i}$ and $f_{\lambda,j}$, their relative optical depth is
    \begin{equation}\label{eq:tau_ij}
      \tau_{i,j}(\lambda) = -\ln\frac{f_{\lambda,i}}{f_{\lambda,j}}.
    \end{equation}
    Then, the selective attenuation can be determined from 
    \begin{equation}\label{eq:Qij}
      Q_{i,j}(\lambda) = \frac{\tau_{i,j}(\lambda)}{\delta\tau_{\mathrm{B},i,j}},
    \end{equation}
    in which $\delta\tau_{\mathrm{B},i,j}$ is the difference between the Balmer optical depths of the two spectra, i.e., $\delta\tau_{\mathrm{B},i,j}=\tau_{\mathrm{B},i}-\tau_{\mathrm{B},j}$ \citep{Calzetti1994}. Finally, the total-to-selective extinction can be expressed as $k(\lambda)=fQ_{i,j}(\lambda)+R_V$ \citep{Battisti2016} if
    \begin{equation}\label{eq:f}
      f = \frac{k_{\mathrm{H\beta}}-k_{\mathrm{H\alpha}}}{E(B-V)_{\mathrm{star}}/E(B-V)_{\mathrm{gas}}}.
    \end{equation}
    Note that the extinction curve of ionized gas used to calculate $k_{\mathrm{H\alpha}}$ and $k_{\mathrm{H\beta}}$ is different from the attenuation curve of the underlying stellar populations that we derive from $Q_{i,j}(\lambda)$. From the above formula, it is clear that the factor $f$ describes the difference between dust attenuation of ionized gas and stars.

    Therefore, the basic steps of this strategy are (1) select spectra that meet the third assumption to the maximum extent; (2) divide them into bins with different stellar reddening according to their \tauB; (3) calculate $Q_{i,j}(\lambda)$ for the average spectra of two bins with different \tauB; and (4) determine $f$ and $R_V$.

  \subsection{Application to IFU Data}
  \label{ssec:application_to_IFU}

    Due to the complication of galactic SFH, the third assumption listed in Section \ref{ssec:basis} is nearly broken for the overall SFG population. To meet this assumption approximately, physical properties related to SFH, such as age indicators (e.g., \dn; \citealt{Battisti2016}) or sSFR \citep{Reddy2015}, are used to select subsamples with similar intrinsic SEDs. In the case of IFU spectra, given the pixel size of $0\farcs5\times0\farcs5$, one spaxel of the MaNGA survey corresponds to a region of $\sim 0.3\times0.3$ kpc at $z=0.03$ that is slightly larger than/comparable to the size of normal star-forming regions (e.g., \citealt{Gusev2014}), while the spatial resolution is about 1.5 kpc (2\farcs5) at the same redshift. Therefore, the SFH of individual spaxels should be much simpler than that of individual galaxies. However, additional age criterion is also needed to obtain a cleaner sample.

\begin{figure}[htb]
    \centering
    \includegraphics[width=0.495\textwidth]{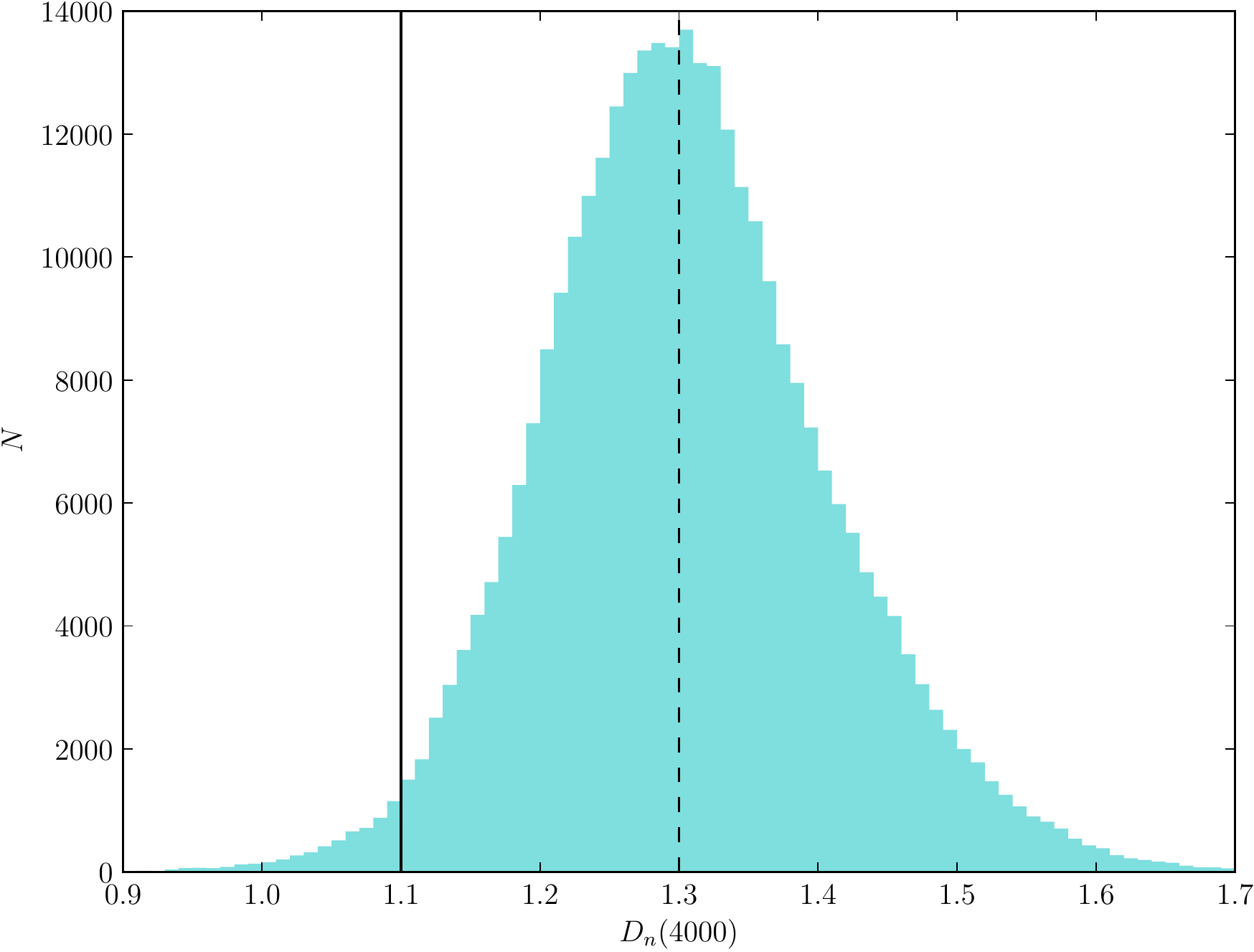}
    \caption{Distribution of \dn\ for all of the star-forming spaxels selected from Section \ref{ssub:selection}. The black solid and dashed lines indicate \dn\ of 1.1 and 1.3, respectively, which enclose the \dn\ window used to construct attenuation curves in this work.}
    \label{fig:hist_dn4000}
    \end{figure}

    As shown in \cite{Kauffmann2003}, both the Balmer absorption line index \hd\ and the 4000 \AA\ spectral break defined by narrow continuum bands (\dn; \citealt{Balogh1999}) are sensitive to the age of stellar population. The strength of \dn\ increase with the stellar population becoming older, while \hd\ peaks at an age of $\sim 5\times10^8$ yr and decreases for both younger and older populations \citep{Kauffmann2003}. In this work, we determine to use \dn\ as age indicator to select spaxels with similar ages due to its positive correlation with age. \cite{Kauffmann2003} also demonstrated that the evolution of \dn\ depends strongly on metallicity when age $> 10^9$ yr. To reduce the effect of metallicity, we constrain our sample to spaxels with $1.1\leq D_n(4000)<1.3$ and divide them into two \dn\ bins with a width of $\Delta D_n(4000)=0.1$ to minimize the effects of stellar population age on the derived attenuation curve. The number of spaxels with $D_n(4000)<1.1$ is only $\sim 6,000$ that is too small to set as a single \dn\ bin. Within the adopted \dn\ window, our final sample consists of about 157,000 spaxels from 982 SFGs, which is nearly half of the star-forming sample selected from Section \ref{ssub:selection}. We present the distribution of \dn\ for all of the star-forming spaxels in Figure \ref{fig:hist_dn4000} in which the \dn\ window of our final sample is also marked out.

    Utilizing the Gaussian fitting results from Section \ref{ssub:data_processing}, we subtract emission lines from the deredshifted spectra to create emission line-free spectra, which are then smoothed using a running median in a wavelength window of 100 \AA. The smoothed spectra are normalized to their rest-frame flux densities at $\lambda=5500$ \AA. Within each \dn\ bin, we divide spaxels into \tauB\ bins with a step of $\Delta\tau_{\mathrm{B}}=0.1$. An average spectrum is obtained for spaxels within the same \tauB\ bin, and is taken as the template spectrum of the bin. Although the observed spectra of the MaNGA survey span a wide wavelength range of 3600--10300 \AA, we restrict our analysis to rest-frame wavelengths of 3800--9000 \AA\ due to the following reasons: (1) our sample covers a redshift range of 0.01--0.15 that limits our analysis to rest-frame wavelengths shorter than $\sim9000$ \AA; and (2) in this work we assume that the observed continua consist of only emission lines and the underlying stellar emission, i.e., we ignore the nebular continua that might have significant contribution to the observed spectra at rest-frame wavelengths of $\lesssim3800$ \AA, especially for those regions with very young stellar populations (e.g., \citealt{Cullen2018}).

    \begin{figure*}[htb]
    \centering
    \includegraphics[width=\textwidth]{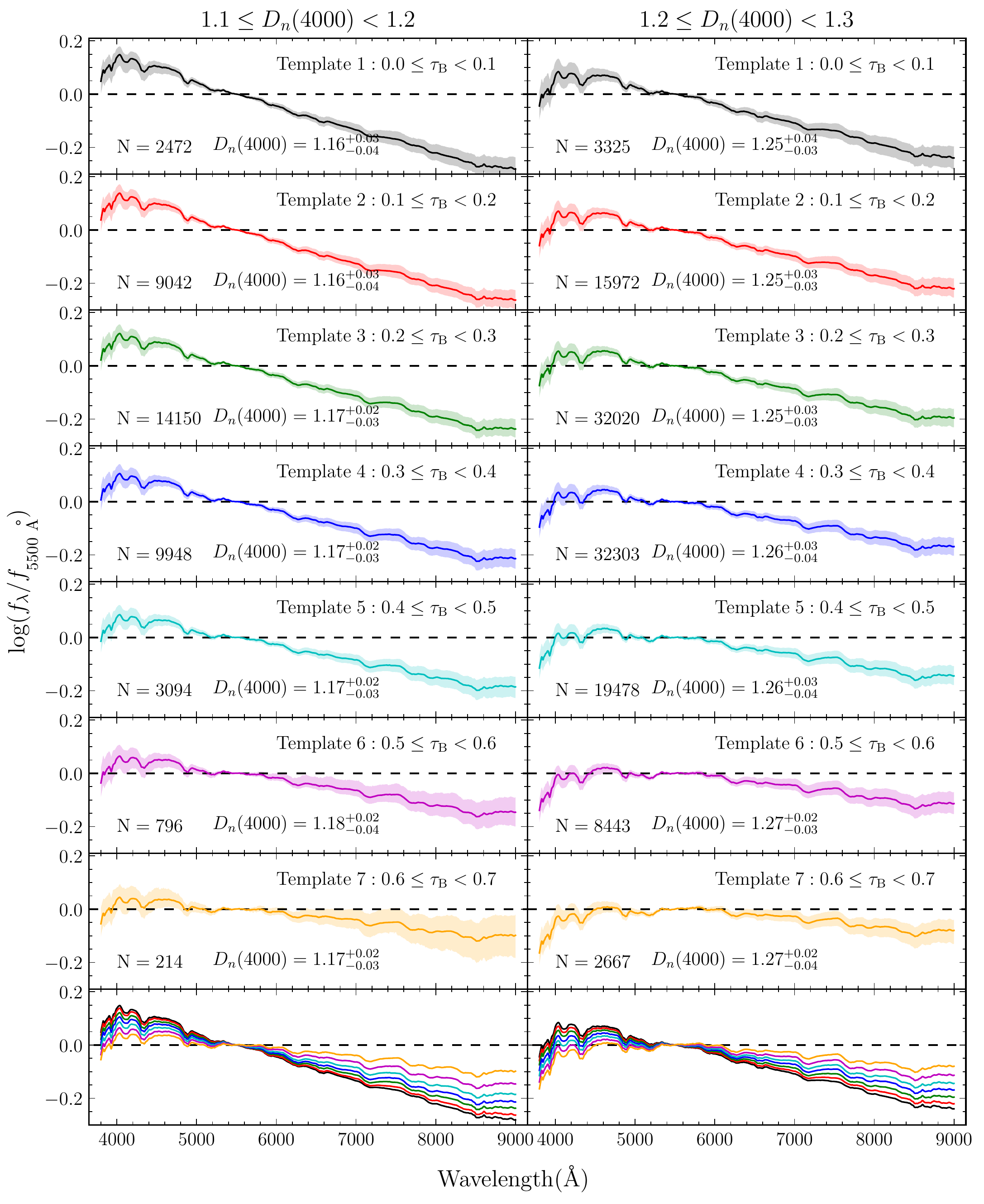}
    \caption{Average spectra in different \tauB\ bins for spaxels with $1.1\leq D_n(4000)<1.2$ (left) and $1.2\leq D_n(4000)<1.3$ (right), normalized at 5500 \AA. The \tauB\ ranges, the numbers of spaxels, and the medians and 68\% ranges of \dn\ of each \tauB\ bin are listed in each panel. The colored shaded area indicate the 68\% scatters around the mean spectra. A direct comparison of all mean spectra is given in the bottom panel for each \dn\ subsample.}
    \label{fig:templates}
    \end{figure*}

    In Figure \ref{fig:templates} we present template spectra for different dust attenuation bins with $0\leq\tau_{\mathrm{B}}<0.7$, together with the colored shadows indicating the 68\% ranges around the mean spectra. The numbers of spaxels and the \dn\ ranges (i.e., 68\% ranges around the median values) within each \tauB\ bin are also listed. The \dn\ ranges of the templates are consistent with each other, suggesting a good approximate for the third assumption presented in Section \ref{ssec:basis}. Directly comparisons of the templates are shown in the bottom panels for each \dn\ bin. There is a clear trend that the shape of templates becomes flatter as the \tauB\ increases, indicating a higher dust reddening of the stellar continua.

    The selective attenuation curves can be derived for each pair of templates via Equation (\ref{eq:Qij}) in which $i$ ($j$) denotes the template with larger (smaller) \tauB. As the dispersions of templates with spaxels numbers less than 500 are fairly large, we do not include these templates in the calculation. Additionally, due to the similar shape of the Template 1 compared with that of Template 2, especially at blue end, we remove Template 1 from our calculation for both \dn\ subsamples to avoid over-weight of Template 1. In \cite{Battisti2016} the template with the smallest \tauB\ was also excluded from their calculation due to similar reason. Therefore, for $1.1\leq D_n(4000)<1.2$, only Templates 2--6 are used, while for $1.2\leq D_n(4000)<1.3$, Templates 2--7 are used. The combinations of these templates will result in 10 and 15 $Q_{i,j}$ for the two \dn\ subsamples, respectively.

\section{Results}
\label{sec:results}
  In this section, we will present the attenuation curve for the selected star-forming regions, and examine how the curve varies with local and global physical properties such as \dn, equivalent widths of \ha\ emission line EW(\ha), inclinations ($b/a$), distances to the galactic center, and so on.

  \subsection{Attenuation Curve of All Spaxels}
  \label{ssec:global}

\begin{figure*}[htb]
    \centering
    \includegraphics[width=\textwidth]{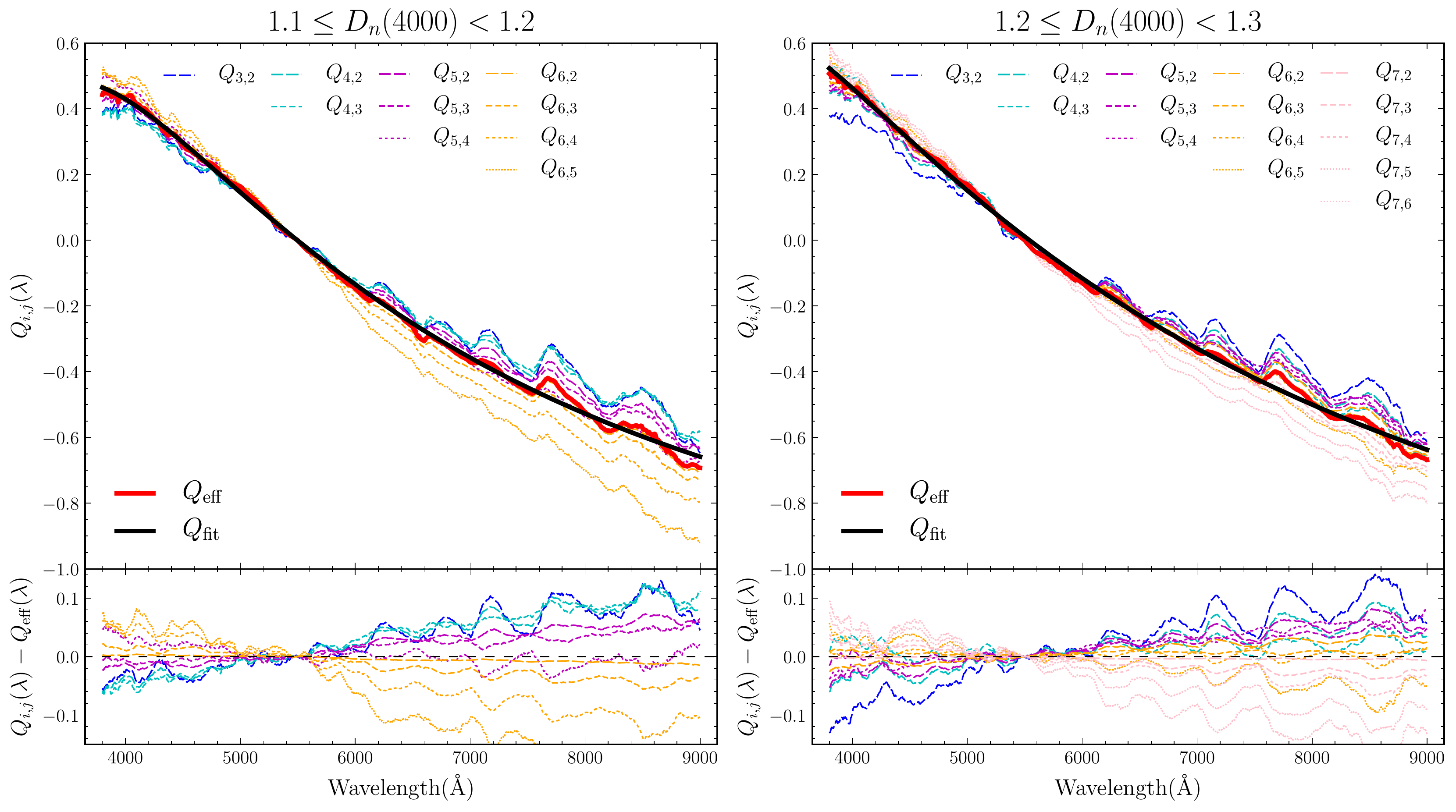}
    \caption{Selective attenuation curve $Q_{i,j}(\lambda)$ derived from all adopted template spectra for $1.1\leq D_n(4000)<1.2$ (left) and $1.2\leq D_n(4000)<1.3$ (right). The colored thin curves indicate $Q_{i,j}(\lambda)$ computed from the comparisons of Template $i$ and Template $j$ presented in Figure \ref{fig:templates}. The solid red line is the effective attenuation curve $Q_{\mathrm{eff}}(\lambda)$ calculated from averaging all $Q_{i,j}(\lambda)$. The black solid line represents the $Q_{\mathrm{fit}}(\lambda)$ that is the best-fit result for $Q_{\mathrm{eff}}(\lambda)$. The differences between $Q_{i,j}(\lambda)$ and $Q_{\mathrm{eff}}(\lambda)$ are shown in the bottom panel for each \dn\ subsample. The differences between $Q_{i,j}(\lambda)$ within the same \dn\ bin reflect variations in the differential reddening between the stellar and ionized components (i.e., the $f$ value).
    \label{fig:Qij}}  
    \end{figure*}

    All of the derived selective attenuation curves are shown in Figure \ref{fig:Qij}. As the templates are normalized at $\lambda = 5500$ \AA, it follows that $Q_{i,j}(5500 \mathrm{\AA})=0$ according to Equation (\ref{eq:Qij}). Following \cite{Battisti2016}, for each \dn\ subsample, an effective attenuation curve, denoted as $Q_{\mathrm{eff}}(\lambda)$, is calculated by averaging all the selective attenuation curve $Q_{i,j}(\lambda)$, and is fitted with a third-order polynomial in form of
    \begin{equation}
      Q_{\mathrm{fit}}(x) = p_0 + p_1 x + p_2 x^2 + p_3 x^3,
    \end{equation}
    where $x=1/\lambda$, and $\lambda$ is in units of \AA. The best-fit results within the wavelength range of 3800--9000 \AA\ are
    \begin{align}
        Q_{\mathrm{fit}}(\lambda) &= -1.3448 + 0.1151x + 0.6238x^2 - 0.1544x^3,\\
        Q_{\mathrm{fit}}(\lambda) &= -1.7998 + 1.0351x + 0.0596x^2 - 0.0447x^3,
    \end{align}
    for $1.1\leq D_n(4000)<1.2$ and $1.2\leq D_n(4000)<1.3$, respectively. $Q_{\mathrm{eff}}(\lambda)$ and $Q_{\mathrm{fit}}(\lambda)$ are also shown in Figure \ref{fig:Qij}.

    As mentioned above, the selective attenuation can be linked to the total-to-select attenuation via $k({\lambda})=fQ(\lambda)+R_V$. By definition, $k(B)-k(V)\equiv 1$. Therefore, in case of $Q_{\mathrm{fit}}(\lambda)$, the factor $f$ can be derived from
    \begin{equation}\label{eq:f_cal}
      f = \frac{1}{Q_{\mathrm{fit}}(B) - Q_{\mathrm{fit}}(V)}.
    \end{equation}
    Here, we assume the wavelength of $B$ and $V$ bands to be 4400 \AA\ and 5500 \AA, respectively. According to this formula, the similar shape of $Q_{\mathrm{fit}}(\lambda)$ exhibited in Figure \ref{fig:Qij} suggests a similar $f$ for two \dn\ subsamples. The resulting $f$ are $3.05_{-0.55}^{+0.50}$ and $3.07_{-0.52}^{+0.79}$ for $1.1\leq D_n(4000)<1.2$ and $1.2\leq D_n(4000)<1.3$, respectively. The upper and lower limits of $f$ denote the ranges of $f$ for individual $Q_{i,j}(\lambda)$.

    Following from Equation (\ref{eq:f}), the large range of $f$ for individual $Q_{i,j}(\lambda)$ within the same \dn\ bin suggests there are significant variations in the differential reddening between the stellar and nebular components. Clearly, a steeper $Q_{i,j}(\lambda)$ (corresponding to a lower $f$) will lead to a larger $E(B-V)_{\mathrm{star}}/E(B-V)_{\mathrm{gas}}$ ratio. Furthermore, it is evident that $Q_{i,j}(\lambda)$ tends to be steeper for a larger ``$i$'' (i.e., the more attenuated templates in Equation (\ref{eq:tau_ij}) have a larger $\tau_{\mathrm{B}}$) within each $D_n(4000)$ bin. This trend implies a smaller difference between the stellar and nebular dust reddening for more obscured stellar/ionized gas regions, given a fixed \dn.

    \begin{figure*}[htb]
    \includegraphics[width=0.5\textwidth]{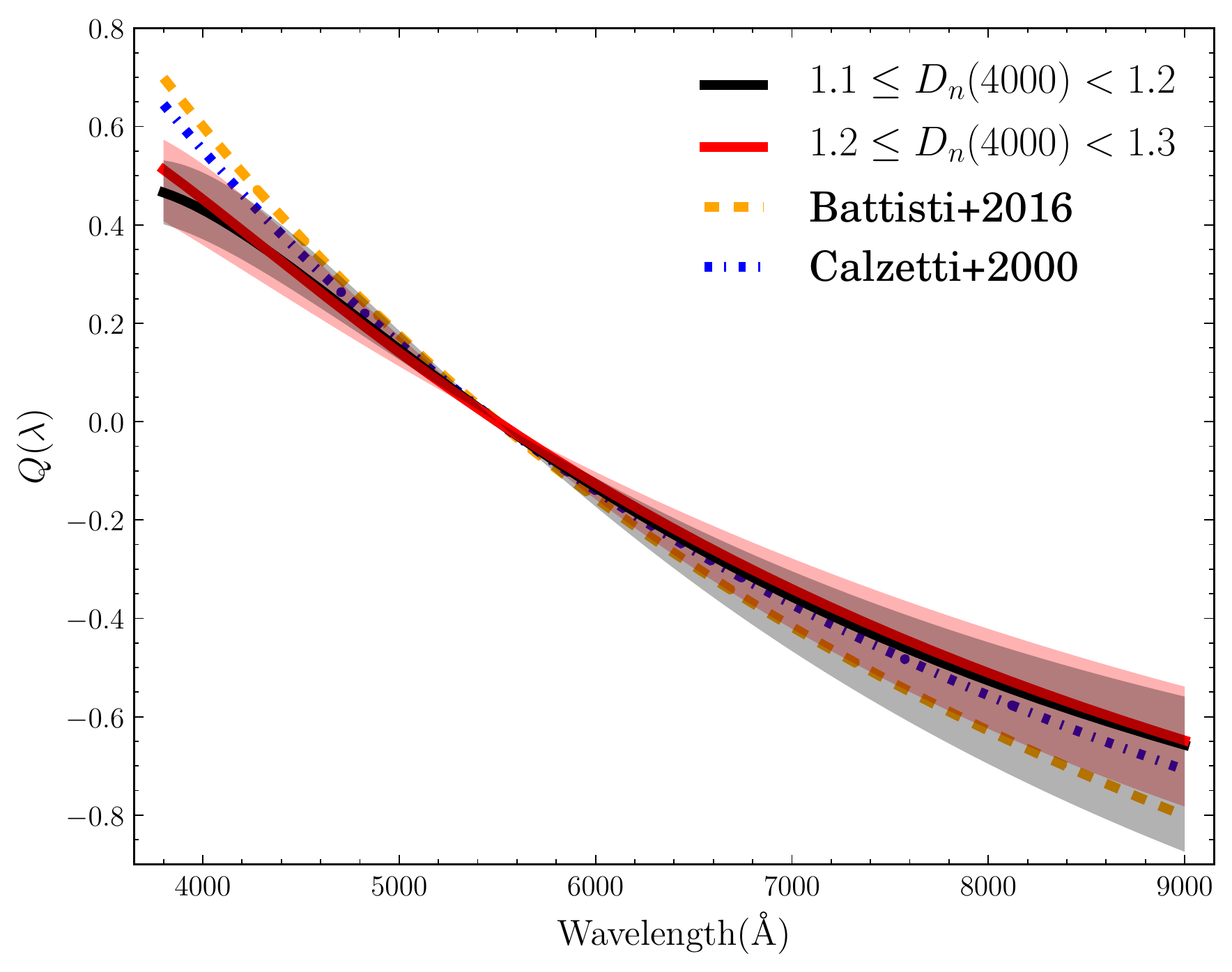}
    \includegraphics[width=0.5\textwidth]{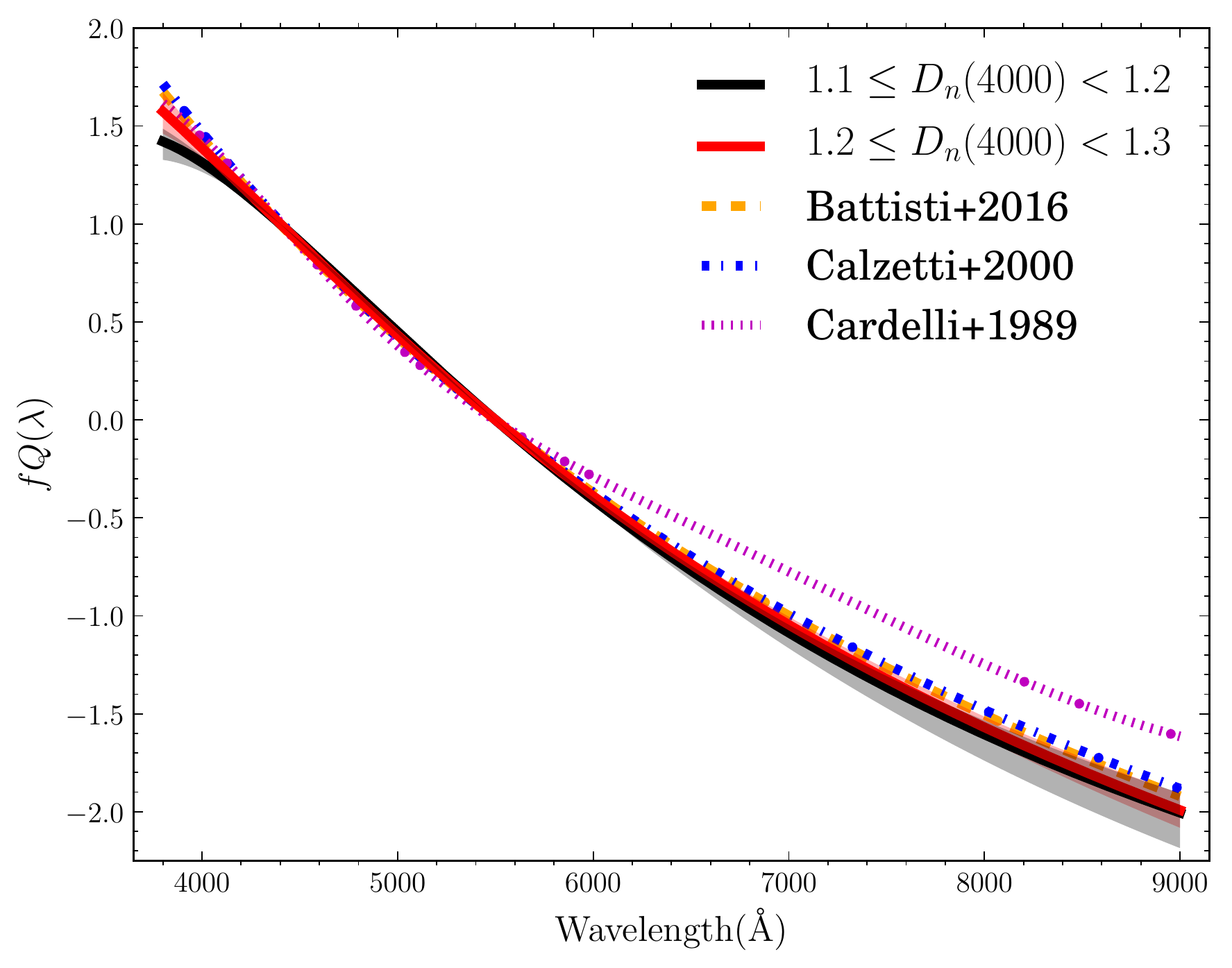}
    \caption{Comparison between our attenuation curves, i.e., $Q(\lambda)$ (left) and $fQ(\lambda)$ (right), and those from previous studies. The black and red solid lines indicate our results for spaxels with $1.1\leq D_n(4000)<1.2$ and $1.2\leq D_n(4000)<1.3$, respectively. The colored shadow regions represent the corresponding area spanned by all $Q_{i,j}(\lambda)$ (in case of $fQ(\lambda)$, $f$ is calculated for each $Q_{i,j}(\lambda)$). The orange dashed line and blue dashed--dotted line denote the results from \cite{Battisti2016} and \cite{Calzetti2000}, respectively. The magenta dotted curve indicates the Milky Way extinction curve of \cite{Cardelli1989} subtracted by $R_V=3.1$. All attenuation/extinction curves are normalized at $\lambda=5500$ \AA.
    \label{fig:q_and_fq}}
    \end{figure*}

    In Figure \ref{fig:q_and_fq}, we compare our results with attenuation curves derived from local starburst galaxies, i.e., the widely used \cite{Calzetti2000} curve, and local SFGs \citep{Battisti2016}. The Milky Way extinction curve of \cite{Cardelli1989} is also shown to compare with $fQ_{\mathrm{fit}}(\lambda)$. The shadow regions indicate the coverage of the fitting results for individual $Q_{i,j}(\lambda)$ (and the corresponding $fQ(\lambda)$) , which can be regarded as the scatters of $Q_{\mathrm{fit}}(\lambda)$ ($fQ_{\mathrm{fit}}(\lambda)$).

    Our selective attenuation curves (both $Q_{\mathrm{fit}}(\lambda)$ and $fQ_{\mathrm{fit}}(\lambda)$) for the two \dn\ subsamples are consistent with each other within the scatters, indicating that the shape of attenuation curve is independent of \dn\ within the wavelength range and \dn\ range we explored in this work. Similar results were also reported in \cite{Battisti2016}, but at galactic scale. Within the scatters, the $Q_{\mathrm{fit}}(\lambda)$ and $fQ_{\mathrm{fit}}(\lambda)$ are roughly consistent with those of \cite{Calzetti2000} and \cite{Battisti2016} at $\lambda>4400$ \AA, regardless of the \dn\ ranges. At the blue end, the $fQ_{\mathrm{fit}}(\lambda)$ curve for the subsample of $1.2\leq D_n(4000)<1.3$ is also in agreement with the reference ones, while a flattening is observed for spaxels with $1.1\leq D_n(4000)<1.2$. By comparing with Figure \ref{fig:Qij}, we note that this flattening arises from the $Q_{i,j}$ with small $i$ values (small Balmer optical depth for the more attenuated template). However, due to the lack of observations at shorter wavelength, we cannot ensure the existence of this feature at present. Observations at UV is needed to confirm the flattening. For this reason, we will focus on the shape of selective attenuation curves at the red end (mainly $\lambda>4400$ \AA) in the following analysis.

    The above results indicate that local subgalactic star-forming regions and SFGs share the same averaged optical attenuation curve in the wavelength range of 3800--9000 \AA. The agreements of the shape of selective attenuation and $f$ between two \dn\ subsamples suggest a similar $E(B-V)_{\mathrm{star}}/E(B-V)_{\mathrm{gas}}$ ratio for them on average according to Equation (\ref{eq:f}), although considerable variations as a function of Balmer optical depth, as well as possible variations on a region-by-region basis, still exist.

  \subsection{Dependence on Local Physical Properties}
  \label{ssec:local_properties}

    Making full use of the IFU data, we first consider whether the shape of dust attenuation curve varies with some local physical properties (i.e., physical properties derived from individual spaxel), such as stellar mass surface density ($\Sigma_*$), SFR surface density ($\Sigma_{\mathrm{SFR}}$), equivalent width of the \ha\ emission line EW(\ha), O3N2 index, and the location of spaxels within the host galaxies. Most of previous studies used the specific star formation rate (sSFR) to characterize the star formation activity (e.g., \citealt{Wild2011,Battisti2016}), however, the EW(\ha) also can be a good proxy of sSFR but is independent of models. Thus, we adopt the EW(\ha) of each spaxel to describe the level of local star formation activity. Note that EW(\ha) used in this work is an absolute value with higher value denoting more intense star formation. The O3N2 index is defined as
    \begin{equation}
      \mathrm{O3N2}\equiv\log\left(\frac{\oiii\lambda5007}{\mathrm{H\beta}}\times\frac{\mathrm{H\alpha}}{\nii\lambda6584}\right),
    \end{equation}
    and can be used as an indicator of gas-phase metallicity (e.g., \citealt{Kewley2008,Marino2013,LinZ2017}). To characterize the location of spaxels, we use the deprojected distances to the galactic centers normalized by the effective radius $r_{\mathrm{e}}$ ($r/r_{\mathrm{e}}$).

    \begin{deluxetable*}{lCCCCCC}
    \tablecaption{Ranges of physical properties for binned samples\label{tab:bin_range}}
    \tablehead{
    \colhead{Parameters} & \multicolumn{3}{c}{$1.1\leq D_n(4000)<1.2$} & \multicolumn{3}{c}{$1.2\leq D_n(4000)<1.3$} \\
    \cline{2-4} \cline{5-7}
    & \colhead{Median} & \colhead{Low-value bin} & \colhead{High-value bin} & \colhead{Median} & \colhead{Low-value bin} & \colhead{High-value bin}
    }
    \startdata
    $\log\Sigma_*(M_{\odot}~{\rm kpc^{-2}})$ & 7.57 & 7.39_{-0.16}^{+0.12} & 7.82_{-0.18}^{+0.31} & 7.81 & 7.60_{-0.19}^{+0.14} & 8.05_{-0.17}^{+0.27} \\
    $\log\Sigma_{\mathrm{SFR}}(M_{\odot}~{\rm yr^{-1}~kpc^{-2}})$ & -1.61 & -1.87_{-0.29}^{+0.18} & -1.32_{-0.20}^{+0.32} & -1.81 & -2.05_{-0.24}^{+0.16} & -1.54_{-0.20}^{+0.35} \\
    $\mathrm{EW(H\alpha)(\AA)}$ & 52.00 & 38.41_{-12.61}^{+9.29} & 69.16_{-12.45}^{+24.48} & 28.13 & 21.75_{-5.55}^{+4.31} & 36.54_{-6.16}^{+12.25} \\
    $\mathrm{O3N2}$ & 0.51 & 0.24_{-0.27}^{+0.18} & 0.81_{-0.22}^{+0.39} & 0.15 & -0.06_{-0.15}^{+0.14} & 0.46_{-0.22}^{+0.38} \\
    $r/r_{\mathrm{e}}$ & 0.82 & 0.55_{-0.26}^{+0.19} & 1.10_{-0.19}^{+0.29} & 0.73 & 0.49_{-0.22}^{+0.16} & 0.98_{-0.18}^{+0.32} \\
    $b/a$ & 0.69 & 0.49_{-0.17}^{+0.13} & 0.82_{-0.08}^{+0.11} & 0.65 & 0.45_{-0.18}^{+0.13} & 0.81_{-0.11}^{+0.11} \\
    $\log M_*(M_{\odot})$ & 9.82 & 9.43_{-0.55}^{+0.25} & 10.09_{-0.16}^{+0.35} & 9.98 & 9.53_{-0.46}^{+0.30} & 10.31_{-0.24}^{+0.32} \\
    \enddata
    \tablecomments{Ranges of physical properties for subsamples considered in this work. For each \dn\ bin and each physical property, the median of the whole subsample, as well as the medians and 68\% dispersions of the binned distributions for the low- and high-value bins, are given.}
    \end{deluxetable*}

    Within each \dn\ subsample, we divide spaxels into two bins according to the median value of each parameter. The medians of these physical properties for each \dn\ subsample, as well as the medians and 68\% dispersions of the binned distributions for the low- and high-value bins, are given in Table \ref{tab:bin_range}.

    We note that although our aim of this subsection is to explore any dependence of the slope of dust attenuation curve on the aforementioned physical properties, a \cite{Calzetti2000} attenuation curve is already adopted during the calculations of these physical properties (mainly $\Sigma_*$, $\Sigma_{\mathrm{SFR}}$, and the O3N2 index) as described in Section \ref{ssub:data_processing}. On the one hand, $\Sigma_*$ is derived from the continuum fitting and is less affected by the assumed attenuation curve, as stellar mass estimation via modeling spectra is not sensitive to the treatment of dust attenuation \citep{Conroy2013}. On the other hand, the reddening correction for emission lines only affects the absolute values of emission line luminosities (and thus $\Sigma_{\mathrm{SFR}}$), but not the relative values of them. The O3N2 index is defined by two ratios of emission line pairs that have very close wavelengths, indicating a negligible reddening correction for this index. Given that our subsamples are generated based on the relative values of these properties, we believe that the assumed attenuation curve should not significantly affect our subsample division, thus the derived attenuation curves.

    \begin{figure*}
    \centering
    \includegraphics[width=0.9\textwidth]{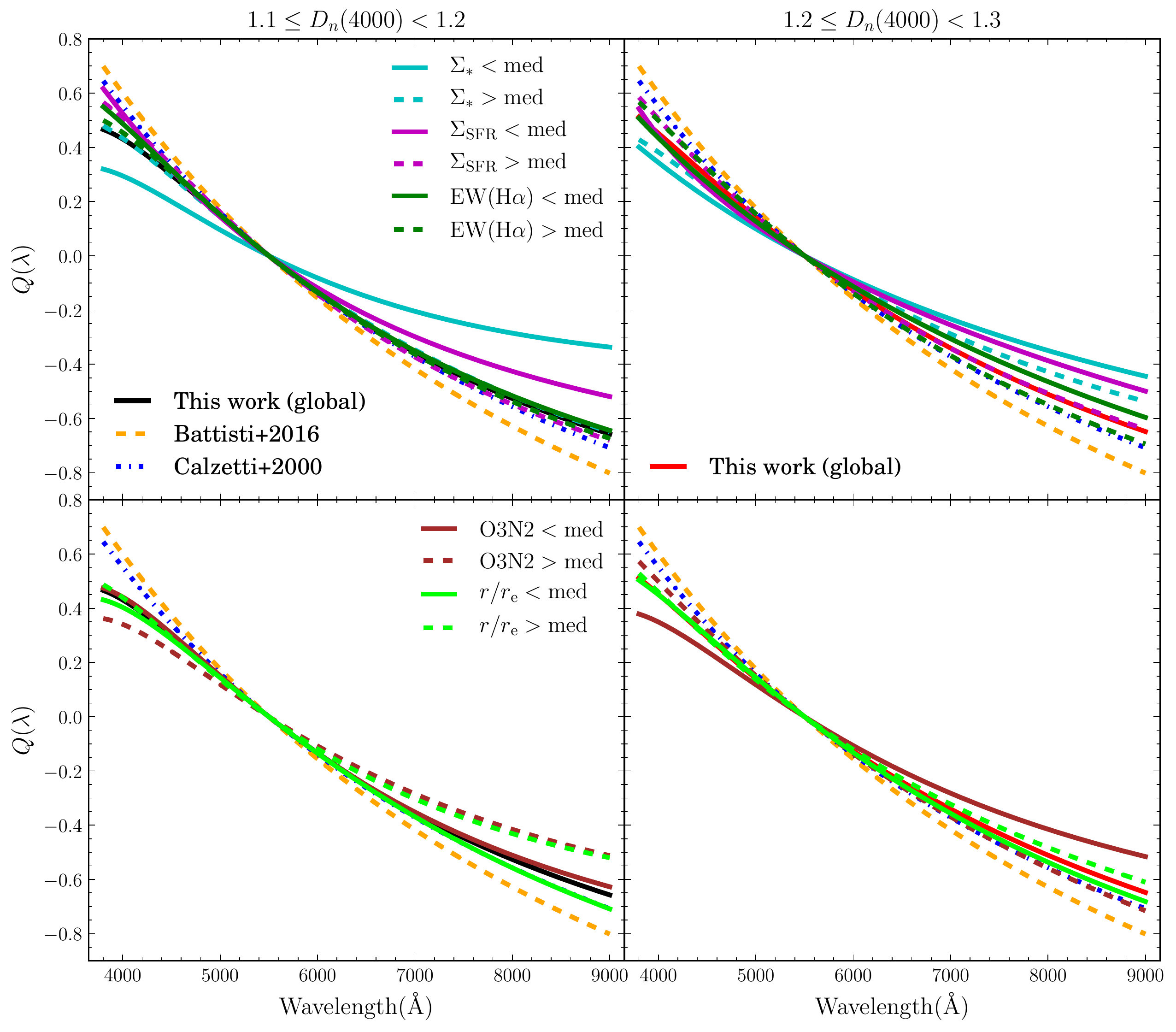}
    \caption{Selective attenuation curve $Q(\lambda)$ in bins of local physical properties ($\Sigma_*$, $\Sigma_{\mathrm{SFR}}$, EW(\ha), O3N2 index, and $r/r_{\mathrm{e}}$) for the $1.1\leq D_n(4000)<1.2$ (left) and $1.2\leq D_n(4000)<1.3$ (right) subsamples. For each physical properties, the solid curves indicate the results of the low-value bins, while the dashed curves are derived from the high-value bins. The black and red solid curves represent the global ones of this work shown in Figure \ref{fig:q_and_fq}. The orange dashed curve and blue dashed--dotted curve are the attenuation curves of \cite{Battisti2016} and \cite{Calzetti2000}, respectively.
    \label{fig:bin_Q_local}}
    \end{figure*}

    \begin{figure*}
    \centering
    \includegraphics[width=0.9\textwidth]{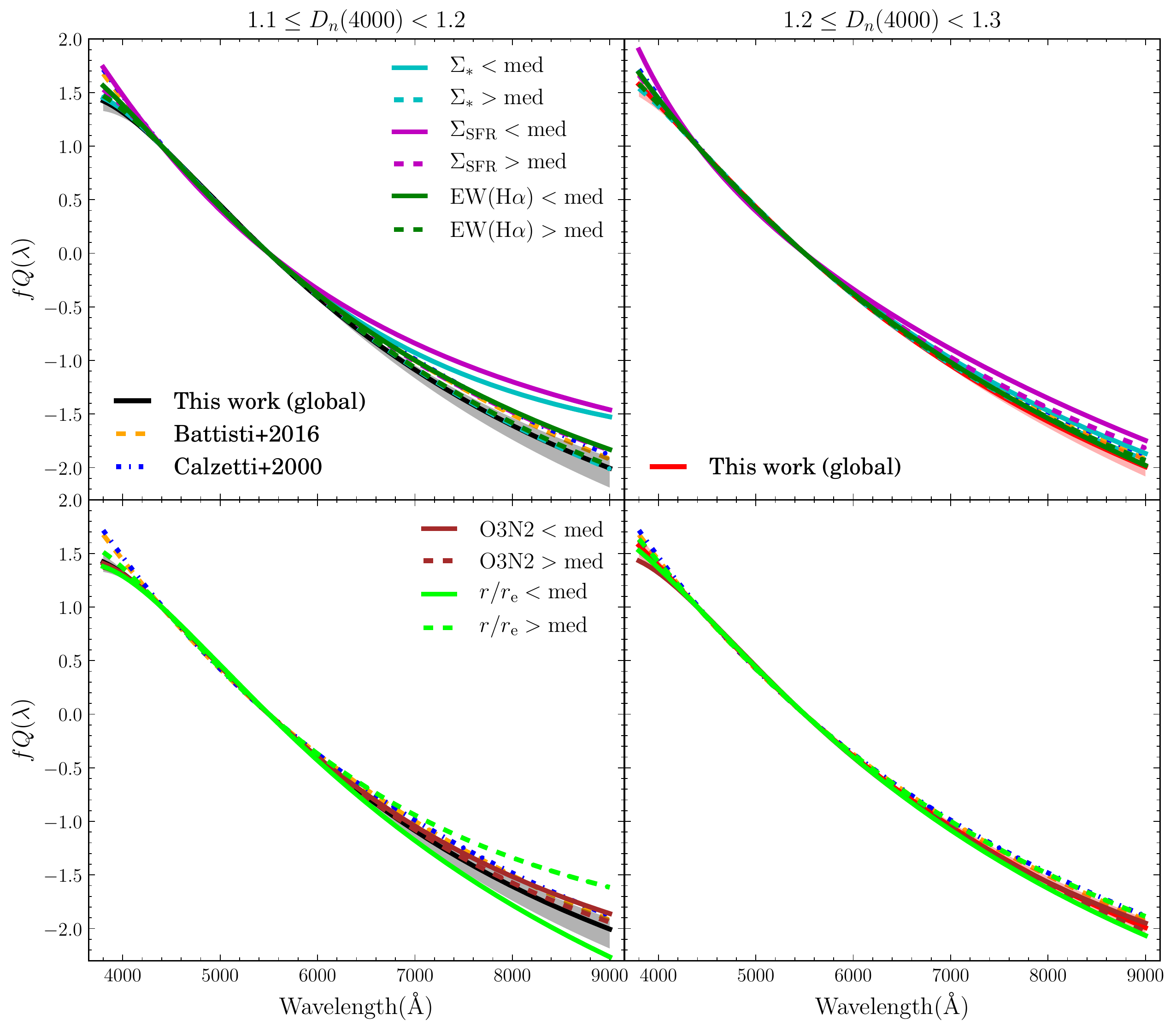}
    \caption{Same as Figure \ref{fig:bin_Q_local} but for selective attenuation curve $fQ(\lambda)$. The colored shadow regions around the global $fQ(\lambda)$ are the same as those in the right panel of Figure \ref{fig:q_and_fq}, representing the corresponding areas spanned by all $fQ_{i,j}(\lambda)$.
    \label{fig:bin_fQ_local}}
    \end{figure*}

    The calculation of the binned selective attenuation curves $Q(\lambda)$ and $fQ(\lambda)$ is the same as the global ones, while the results for the two \dn\ subsamples are shown in Figures \ref{fig:bin_Q_local} and \ref{fig:bin_fQ_local}, respectively, comparing with the ones for the overall \dn\ subsamples plotted in Figure \ref{fig:q_and_fq} (labeled as ``global'') and two reference attenuation curves from \cite{Calzetti2000} and \cite{Battisti2016}. The best-fit parameters of the selective attenuation curves for all binned samples are listed in Table \ref{tab:p_bin}. The diversity of the shapes of $Q(\lambda)$ for the binned subsamples reveal remarkable variation in the $f$ factor, and thus in $E(B-V)_{\mathrm{star}}/E(B-V)_{\mathrm{gas}}$. 

    For the $fQ(\lambda)$, the bins with $\Sigma_*<$ median and $\Sigma_{\mathrm{SFR}}<$ median exhibit large deviations at longer wavelengths from the global one, as well as the referenced starbursts and normal SFG attenuation curves when $1.1\leq D_n(4000)<1.2$, suggesting a flatter curve for these two bins. As these two properties form a well-studied subgalactic main sequence (SGMS; \citealt{Hsieh2017,Liu2018}), the above results imply a flatter selective attenuation curve for star-forming regions at the low-mass end of the SGMS. In the case of $1.2\leq D_n(4000)<1.3$, the variations in the binned $fQ(\lambda)$ is smaller. For each local physical property, the $fQ(\lambda)$ derived from the two bins are consistent with each other, as well as the global one and two reference attenuation curves, if the scatters are taken into account. Binning by the average stellar mass surface density ($\mu_*$) of galaxies, \cite{Wild2011} also reported a steeper optical slope for their high-$\mu_*$ sample compared with the low-$\mu_*$ sample, which is consistent with our result of the younger subsample. However, our results further imply an age-dependence of this behavior that was not explored in \cite{Wild2011}.

    Intriguingly, we find that the binning in $r/r_{\mathrm{e}}$ also gives significant difference in the $fQ(\lambda)$ for $1.1\leq D_n(4000)<1.2$. With the \dn\ range, spaxels in the inner regions of galaxies have a steeper selective attenuation at longer wavelengths compared with the global curve, while outer regions present a much flatter selective attenuation. However, this difference vanishes for the $1.2\leq D_n(4000)<1.3$ subsample. The location of spaxel is not a physical parameter, there should be other more intrinsic physical properties related to the location to account for the observation. Given the similar behavior of the $\Sigma_*<$ median and $\Sigma_{\mathrm{SFR}}<$ median subsamples to the $r/r_{\rm e} >$ median subsample in $fQ(\lambda)$, we argue that these two surface densities might be such more intrinsic properties. On the one hand, the outer regions of galaxies tend to have smaller $\Sigma_*$ and $\Sigma_{\mathrm{SFR}}$ compared with the inner regions. On the other hand, we have demonstrated that spaxels with smaller $\Sigma_*$ and $\Sigma_{\mathrm{SFR}}$ give a flatter selective attenuation that is consistent with the $fQ(\lambda)$ of the outer bin. Hence, we speculate that the differences in $\Sigma_*$ and $\Sigma_{\mathrm{SFR}}$ can (at least partly) explain the observed different slopes of the selective attenuation curve between the inner and outer bins.

    Therefore, we conclude that star-forming regions with different sSFR or gas-phase metallicity within the ranges we examined in this work (i.e., Table \ref{tab:bin_range}) share similar average selective attenuation curve, irrespective of the age of stellar population. At the galactic scale, the dependence of the shape/slope of attenuation curve on sSFR is found to be negligible for either local \citep{Battisti2016} or high-redshift \citep{Reddy2015} SFGs. However, for the younger subsample ($1.1\leq D_n(4000)<1.2$), star-forming regions with smaller $\Sigma_*$, smaller $\Sigma_{\mathrm{SFR}}$, or larger $r/r_{\mathrm{e}}$ tend to have a flatter attenuation curves compared with the the global one and two reference curves, while for the slightly older stellar population, no evident difference is observed between the binned results for all local properties we explored. These results indicate a different size distribution of dust grains (i.e., different underlying extinction curves) or stars/dust geometry (different behavior in scatting; \citealt{Narayanan2018}) between the younger and slightly older stellar populations.

    \begin{splitdeluxetable*}{lCCCCCBlCCCCC}
    \tablecaption{Best-fit Results of Selective Attenuation Curves for Binned Samples\label{tab:p_bin}}
    \tablehead{
    \colhead{Bin} & \multicolumn{5}{c}{$1.1\leq D_n(4000)<1.2$} & \colhead{Bin} & \multicolumn{5}{c}{$1.2\leq D_n(4000)<1.3$} \\
    \cline{2-6} \cline{8-12}
    & \colhead{$p_0$} & \colhead{$p_1$} & \colhead{$p_2$} & \colhead{$p_3$} & \colhead{$f$} & & \colhead{$p_0$} & \colhead{$p_1$} & \colhead{$p_2$} & \colhead{$p_3$} & \colhead{$f$} 
    }
    \startdata
    $\Sigma_*<\mathrm{med}$ & -0.056\pm0.043 & -1.065\pm0.075 & 0.918\pm0.008 & -0.175\pm0.008 & 4.53_{-0.68}^{+0.74} &
    $\Sigma_*<\mathrm{med}$ & -1.334\pm0.034 & 0.930\pm0.060 & -0.112\pm0.034 & 0.004\pm0.006 & 4.20_{-0.44}^{+0.39} \\
    $\Sigma_*>\mathrm{med}$ & -1.539\pm0.040 & 0.522\pm0.069 & 0.373\pm0.007 & -0.106\pm0.007 & 3.10_{-0.70}^{+0.58} &
    $\Sigma_*>\mathrm{med}$ & -1.281\pm0.036 & 0.481\pm0.063 & 0.263\pm0.036 & -0.075\pm0.007 & 3.59_{-0.95}^{+2.83} \\
    $\Sigma_{\mathrm{SFR}}<\mathrm{med}$ & -0.923\pm0.067 & -0.002\pm0.117 & 0.410\pm0.012 & -0.071\pm0.012 & 2.81_{-0.07}^{+0.07} &
    $\Sigma_{\mathrm{SFR}}<\mathrm{med}$ & -1.926\pm0.037 & 1.884\pm0.065 & -0.656\pm0.037 & 0.113\pm0.007 & 3.50_{-0.83}^{+3.31} \\
    $\Sigma_{\mathrm{SFR}}>\mathrm{med}$ & -1.388\pm0.044 & 0.169\pm0.078 & 0.572\pm0.008 & -0.134\pm0.008 & 2.70_{-0.30}^{+0.61} &
    $\Sigma_{\mathrm{SFR}}>\mathrm{med}$ & -1.727\pm0.042 & 0.987\pm0.074 & 0.036\pm0.042 & -0.028\pm0.008 & 2.84_{-0.27}^{+1.67} \\
    $\mathrm{EW(H\alpha)<med}$ & -1.414\pm0.049 & 0.363\pm0.086 & 0.417\pm0.009 & -0.103\pm0.009 & 2.84_{-0.32}^{+0.27} &
    $\mathrm{EW(H\alpha)<med}$ & -1.975\pm0.034 & 1.559\pm0.060 & -0.312\pm0.034 & 0.030\pm0.006 & 3.32_{-0.43}^{+0.77} \\
    $\mathrm{EW(H\alpha)>med}$ & -1.494\pm0.043 & 0.350\pm0.076 & 0.495\pm0.008 & -0.129\pm0.008 & 2.94_{-0.77}^{+0.91} &
    $\mathrm{EW(H\alpha)>med}$ & -1.829\pm0.042 & 0.941\pm0.073 & 0.149\pm0.041 & -0.060\pm0.008 & 2.79_{-0.46}^{+1.16} \\
    $\mathrm{O3N2<med}$ & -0.912\pm0.030 & -0.557\pm0.053 & 0.960\pm0.005 & -0.209\pm0.005 & 2.97_{-0.80}^{+1.31} &
    $\mathrm{O3N2<med}$ & -0.975\pm0.027 & -0.035\pm0.047 & 0.554\pm0.027 & -0.131\pm0.005 & 3.78_{-1.05}^{+8.73} \\
    $\mathrm{O3N2>med}$ & -0.805\pm0.039 & -0.369\pm0.069 & 0.749\pm0.007 & -0.168\pm0.007 & 3.79_{-0.61}^{+0.84} &
    $\mathrm{O3N2>med}$ & -2.249\pm0.034 & 1.626\pm0.060 & -0.214\pm0.034 & 0.002\pm0.006 & 2.84_{-0.62}^{+1.48} \\
    $r/r_{\mathrm{e}}<\mathrm{med}$ & -1.845\pm0.048 & 0.811\pm0.084 & 0.319\pm0.009 & -0.113\pm0.009 & 3.19_{-0.59}^{+0.56} &
    $r/r_{\mathrm{e}}<\mathrm{med}$ & -1.868\pm0.044 & 1.023\pm0.078 & 0.119\pm0.044 & -0.062\pm0.008 & 3.03_{-0.31}^{+0.74} \\
    $r/r_{\mathrm{e}}>\mathrm{med}$ & -0.529\pm0.038 & -0.811\pm0.067 & 0.945\pm0.007 & -0.186\pm0.007 & 3.10_{-0.43}^{+0.25} &
    $r/r_{\mathrm{e}}>\mathrm{med}$ & -1.696\pm0.031 & 1.005\pm0.055 & 0.014\pm0.031 & -0.028\pm0.006 & 3.09_{-0.52}^{+0.50} \\
    \cline{1-12}
    $b/a<\mathrm{med}$ & -1.444\pm0.042 & 0.205\pm0.073 & 0.604\pm0.008 & -0.154\pm0.008 & 2.99_{-0.70}^{+0.84} &
    $b/a<\mathrm{med}$ & -2.069\pm0.040 & 1.381\pm0.071 & -0.100\pm0.040 & -0.017\pm0.007 & 2.89_{-0.34}^{+0.30} \\
    $b/a>\mathrm{med}$ & -1.081\pm0.052 & -0.018\pm0.091 & 0.588\pm0.009 & -0.137\pm0.009 & 3.33_{-0.48}^{+0.63} &
    $b/a>\mathrm{med}$ & -1.091\pm0.040 & 0.192\pm0.070 & 0.420\pm0.039 & -0.106\pm0.007 & 3.82_{-0.32}^{+0.92} \\
    $\log M_*<\mathrm{med}$ & -1.647\pm0.043 & 0.490\pm0.076 & 0.459\pm0.008 & -0.125\pm0.008 & 2.77_{-0.53}^{+0.44} &
    $\log M_*<\mathrm{med}$ & -2.248\pm0.040 & 1.443\pm0.071 & -0.054\pm0.040 & -0.031\pm0.007 & 2.62_{-0.20}^{+0.86} \\
    $\log M_*>\mathrm{med}$ & -0.672\pm0.028 & -0.781\pm0.049 & 1.011\pm0.005 & -0.209\pm0.005 & 3.07_{-0.23}^{+0.31} &
    $\log M_*>\mathrm{med}$ & -1.023\pm0.025 & 0.078\pm0.043 & 0.468\pm0.025 & -0.110\pm0.005 & 3.61_{-1.05}^{+5.87} \\
    \enddata
    \tablecomments{The central values of $f$ listed above are calculated from the $Q_{\mathrm{fit}}(\lambda)$, while the upper and lower limits denote the range of $f$ for individual $Q_{i,j}(\lambda)$ in each case.}
    \end{splitdeluxetable*}

  \subsection{Dependence on Galactic Global Properties}
  \label{ssec:global_properties}

    \begin{figure*}[htb]
    \includegraphics[width=\textwidth]{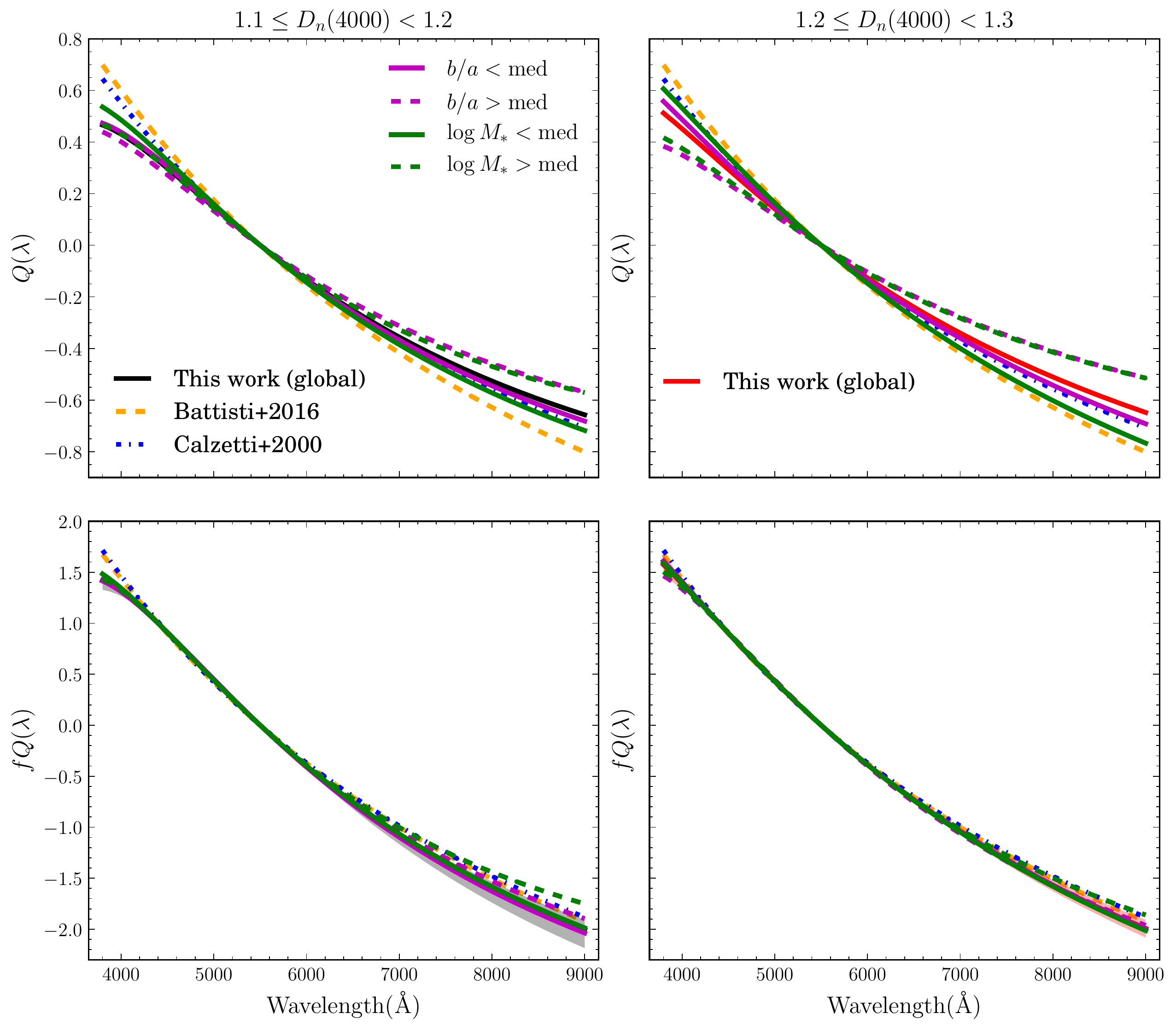}
    \caption{Selective attenuation curve $Q(\lambda)$ (top) and $fQ(\lambda)$ (bottom) in bins of global properties ($b/a$ and $M_*$) for the $1.1\leq D_n(4000)<1.2$ (left) and $1.2\leq D_n(4000)<1.3$ (right) subsamples. For each physical properties, the solid curves indicate the results of the low-value bins, while the dashed curves are derived from the high-value bins. The black and red solid curves represent the global ones of this work shown in Figure \ref{fig:q_and_fq}, while the colored shadow regions around the global $fQ(\lambda)$ are the same as those in the right panel of Figure \ref{fig:q_and_fq}. The orange dashed curve and blue dashed--dotted curve are the attenuation curves of \cite{Battisti2016} and \cite{Calzetti2000}, respectively.
    \label{fig:bin_global}}
    \end{figure*}

    Moving toward properties at larger scale, we examine the dependence of the selective attenuation curve on the axial ratio $b/a$ (i.e., inclination) and the global stellar mass $M_*$ of the host galaxies. Similar to the local cases, the median values of these parameters are used to separate each \dn\ subsample into two bins. Repeating the derivation of the selective attenuation curve, we obtain the $Q(\lambda)$ and $fQ(\lambda)$ for each bin, and show them in Figure \ref{fig:bin_global}. Again, we present the ranges of physical properties and the corresponding best-fit parameters of the selective attenuation curves for all binned samples in Tables \ref{tab:bin_range} and \ref{tab:p_bin}, respectively.

    As we know, stellar light in more edge-on galaxies suffers more reddening (e.g., \citealt{Yip2010}) compared with face-on galaxies due to the larger dust column density. Thus, we would wonder if the inclination has effect on the shape of attenuation curve or not. The $fQ(\lambda)$ in Figure \ref{fig:bin_global} reveal that the inclination ($b/a$) of the host galaxies has negligible effect on the selective attenuation, regardless of the \dn. Based on photometric and spectroscopic data of local galaxies, \cite{Wild2011} found that the slope of attenuation curve in the optical wavelengths increases with increasing $b/a$ ratio. In other words, more face-on galaxies have steeper attenuation curve, which is contrary to our result. However, \cite{Battisti2017a} elucidated that the dependence of the shape of attenuation curve on inclination is mainly at UV wavelengths, whereas the longer region remains unchanged. One should note that our derived attenuation curves focus on the red end of optical wavelength due to the lack of UV observation, thus the negligible effect of inclination on the curve are in good agreement with \cite{Battisti2017a}. \cite{Battisti2017a} also discussed several possible reasons for the different dependence of the slope of attenuation curve on inclination at longer wavelengths between their results and those of \cite{Wild2011}. We here highlight the differences in the adopted aperture. \cite{Wild2011} used photometry from the entire galaxy, while the SDSS spectra used in \cite{Battisti2017a} only enclose the most central regions of galaxies. Similarly, limited by the field of view of the MaNGA survey, the IFU observations only reach out to 1.5$r_{\mathrm{e}}$ for most of the MaNGA targets (i.e., the ``Primary+'' sample; \citealt{Bundy2015}). Thus, our selected spaxels cannot cover the most outer regions of the targeted galaxies. As discussed in \cite{Battisti2017a}, the central and most outer regions of galaxies might have different dust properties, which is already suggested by our result that the shapes of the attenuation curves of star-forming regions located in the inner and outer regions are different.

    For the binning in the global $M_*$, the $fQ(\lambda)$ of the more massive galaxies is slightly flatter than that of the less-massive galaxies for the younger population. However, such differences are not significant. Although all $Q_{i,j}(\lambda)$ of the less-massive bin ($\log M_*<\mathrm{med}$) are steeper than the average curve of the more massive bin, the spanned area of all $Q_{i,j}(\lambda)$ of $\log M_*>\mathrm{med}$ forms a large region and cover the average curve of the less-massive bin, suggesting a large diversity of the shape of dust attenuation curve in more massive galaxies. For the older population, it seems that the shape of the selective attenuation shows no dependence on $M_*$.

    In short, there is no significant evidence of variation with global properties from either $D_n(4000)$ subsample.

\section{Discussion}
\label{sec:discussion}

\subsection{Effect of Stars/Dust Geometry}
\label{sec:assumed_geom}

  As we elucidated in Section \ref{sec:methods}, the derivation of selective attenuation curve assumes that dust attenuation is dominated by a uniform foreground-like dust component. This dust geometry model is consistent with the linear relation between $\tau_B$ and the UV slope ($\beta$) observed in local starburst galaxies \citep{Calzetti1994} or normal SFGs \citep{Battisti2016}, as well as in $z\sim2$ SFGs \citep{Reddy2015}, and enables a straightforward derivation of selective attenuation curve. However, the real stars/dust geometry of star-forming regions might be more complex.

  Based on high spatial resolution observations of the starburst galaxy M83, \cite{Liu2013b} found a large diversity of the dust geometry for star-forming regions at a spatial resolution of 6 pc. Nevertheless, for star-forming regions smoothed to a large physical scale (100--200 pc), the dust attenuation is in good agreement with a uniform dust screen model. In other words, the simple uniform foreground screen assumption of dust geometry should be reasonable for observations with a spatial resolution coarser than $\sim 200$ pc. On the other hand, the reconstructed PSF of the MaNGA data has an FWHM of $\sim$2\farcs5 \citep{Law2016}, which corresponds to a physical size of $\sim1.5$ kpc at the mean redshift of the MaNGA survey \citep{Bundy2015}. Therefore, although the stars/dust geometry of star-forming regions has a large diversity on small physical scales \citep{Liu2013b}, the adopted assumption in this work is plausible for our sample.

  Furthermore, even for more complex dust geometry, the derivation of selective attenuation curve is also applicative. \cite{Calzetti1994} introduced five models for the stars/dust geometry to understand the observed linear $\tau_B$--$\beta$ relation in local starburst galaxies, including
  \begin{enumerate}
    \item a uniform foreground screen model in which dust is uniformly distributed and acts as a screen between the radiation source and the observers;
    \item a uniform scattering slab model in which dust is still uniformly distributed but is located close to the radiation source so that scattering into to the line of sight should be considered;
    \item a clumpy foreground screen model that is similar to the first model but dust is clumpy;
    \item a scattering slab model that is similar to the second model but dust is clumpy; and
    \item an internal dust model in which dust and radiation source are uniformly mixed.
  \end{enumerate}

  We refer the readers to \cite{Calzetti1994} for a detailed description of these models. Note that only the uniform foreground screen model naturally gives Equation (\ref{eq:fobs}) in the sense that $\tau_{\lambda}$, and thus $\tau_B$ and $\tau_{i,j}$ in Equation (\ref{eq:tau_ij}), only depends on dust properties. For the other four models, the expressions between the observed spectra $f_{\lambda,\mathrm{obs}}$ and the intrinsic spectra $f_{\lambda,\mathrm{int}}$ as a function of $\tau_{\lambda}$ are more complex. However, it is also convenient to define an effective optical depth $\tau_{\lambda,\mathrm{eff}}\equiv -\ln(f_{\lambda,\mathrm{obs}}/f_{\lambda,\mathrm{int}})$, which depends on both dust properties and stars/dust geometry \citep{Calzetti1994}. By doing so, the derivation of attenuation curve given in in Section \ref{sec:methods} is still reasonable in term of $\tau_{\lambda,\mathrm{eff}}$.

\subsection{Effect of Diffuse Ionized Gas}
\label{sec:DIG}

  In this work, we simply attribute non-AGN ionized gas to star-forming regions. However, such ionized gas also might have a nonnegligible contribution from DIG, which might be excited by mechanisms other than star formation, such as hot, low-mass evolved stars \citep{Flores-Fajardo2011,Zhang2017} or shock (see \citealt{Haffner2009} for a review), and have a significant contribution to a sample of star-forming regions selected from the MaNGA survey \citep{Lin2020}. DIG-dominated spaxels might not support the assumptions adopted in this work. For instance, an intrinsic \ha/\hb\ ratio of 2.86 in the definition of $\tau_B$ might be not suitable for DIG that does not leak from star-forming regions. Moreover, a positive correlation between stellar and nebular attenuations (i.e., the second assumption listed in Section \ref{ssec:basis}) enables us to use $\tau_B$ as a tracer of stellar attenuation, however, such correlation almost disappears for DIG-dominated spaxels \citep{Lin2020}.

  Fortunately, the DIG-dominated spaxels in the MaNGA survey can be selected by \ha\ luminosity surface brightness (\sigha; \citealt{Zhang2017}) or \ewha\ \citep{Lacerda2018} criterion, i.e., $\Sigma_{\rm H\alpha}<10^{39}~{\rm erg~s^{-1}~kpc^{-2}}$ and ${\rm EW_{H\alpha}<3}$ \AA, respectively. We have checked the final sample that we use to derive the dust attenuation curves, and find that the medians and 68\% ranges of $\log~\Sigma_{\rm H\alpha}/{\rm erg~s^{-1}~kpc^{-2}}$ and \ewha\ are $39.6_{-0.4}^{+0.4}$ and $32_{-11}^{+21}$ \AA, respectively. The fraction of spaxels in our sample that are classified as DIG by the \sigha\ and \ewha\ criteria are only 4.8\% and $<0.1\%$, respectively. Therefore, although DIG-dominated regions might not follow the assumptions listed in Section \ref{ssec:basis} and the assumed intrinsic \ha/\hb\ ratio, the effect arisen from DIG is negligible due to the small fraction of DIG-dominated regions in our sample.

\section{SUMMARY}
\label{sec:summary}

  In this paper, we construct a sample of 157,000 spaxels from 982 SFGs based on the DR14 of the IFS survey of MaNGA. Applying the method described in \cite{Calzetti1994} and \cite{Battisti2016}, we determine the average selective dust attenuation for spatially resolved star-forming regions with $1.1\leq D_n(4000)<1.2$ and $1.2\leq D_n(4000)<1.3$. Further analyses are performed to see whether the shape of optical attenuation curve varies with local and global physical properties. The main results of this work are listed below.

  \begin{itemize}
    \item The average attenuation curve of subgalactic star-forming regions in the wavelength range of 3800--9000 \AA\ shows no dependence on $D_n(4000)$ within $1.1\leq D_n(4000)<1.3$, and is similar to the one derived from either starbursts \citep{Calzetti2000} or normal SFGs \citep{Battisti2016}.
    \item For the younger population ($1.1\leq D_n(4000)<1.2$), spaxels at the low-mass end of the SGMS or resided in the outer regions of the host galaxies tend to have flatter attenuation curves compared with the global ones, while spaxels with different sSFR or gas-phase metallicity show similar attenuation curves. For the older population (i.e., $1.2\leq D_n(4000)<1.3$), no dependence on local physical properties is found for the average attenuation curves. These results may suggest a different size distribution of dust grains or stars/dust geometry between the younger and slightly older stellar populations.
    \item No significant trend with global properties (inclination and $M_*$) in the shape of the derived attenuation curves are observed from either $D_n(4000)$ subsample.
  \end{itemize}

  The above results suggest a diversity of the shape of dust attenuation curve in the optical wavelength for subgalactic star-forming regions. For the relatively old population, there is a fairly uniform dust attenuation curve that is similar to the one of either starbursts \citep{Calzetti2000} or normal SFGs \citep{Battisti2016}. Conversely, remarkable dependences on local physical properties are observed for the younger population, highlighting the risk of applying a single attenuation curve across the disks of SFGs. More detailed analysis on these dependences are restricted by the limited number of high-quality spectra. With the future release of the MaNGA data, we expect to have a better understanding of the behavior of dust attenuation curve.

\acknowledgments
The authors are grateful to the anonymous referee for helpful suggestions.
This work is supported by the National Key R\&D Program of China (2017YFA0402600), the B-type Strategic Priority Program of the Chinese Academy of Sciences (XDB41000000) and the National Natural Science Foundation of China (NSFC, Nos. 11421303 and 11973038). Z. Lin gratefully acknowledges support from the China Scholarship Council (No. 201806340211).

Funding for the Sloan Digital Sky Survey IV has been provided by the Alfred P. Sloan Foundation, the U.S. Department of Energy Office of Science, and the Participating Institutions. SDSS-IV acknowledges support and resources from the Center for High-Performance Computing at the University of Utah. The SDSS website is www.sdss.org. SDSS-IV is managed by the Astrophysical Research Consortium for the Participating Institutions of the SDSS Collaboration including the Brazilian Participation Group, the Carnegie Institution for Science, Carnegie Mellon University, the Chilean Participation Group, the French Participation Group, Harvard-Smithsonian Center for Astrophysics, Instituto de Astrofsica de Canarias, The Johns Hopkins University, Kavli Institute for the Physics and Mathematics of the Universe (IPMU)/University of Tokyo, Lawrence Berkeley National Laboratory, Leibniz Institut fr Astrophysik Potsdam (AIP), Max-Planck-Institut fr Astronomie (MPIA Heidelberg), Max-Planck-Institut fr Astrophysik (MPA Garching), Max-Planck-Institut fr Extraterrestrische Physik (MPE), National Astronomcal Observatory of China, New Mexico State University, New York University, University of Notre Dame, Observatao Nacional/MCTI, The Ohio State University, Pennsylvania State University, Shanghai Astronomical Observatory, United Kingdom Participation Group, Universidad Nacional Autonoma de Mexico, University of Arizona, University of Colorado Boulder, University of Oxford, University of Portsmouth, University of Utah, University of Virginia, University of Washington, University of Wisconsin, Vanderbilt University, and Yale University.

\bibliography{sam}

\begin{thebibliography}{}
\providecommand\natexlab[1]{#1}
\providecommand\JournalTitle[1]{#1}

\bibitem[{Abolfathi {et~al.}(2018)Abolfathi, Aguado, Aguilar, Allende~Prieto,
  Almeida, Tasnim~Ananna, Anders, Anderson, Andrews, Anguiano,
  Arag{\'o}n-Salamanca, Argudo-Fern{\'a}ndez, Armengaud, Ata, Aubourg,
  Avila-Reese, Badenes, Bailey, Balland, Barger, Barrera-Ballesteros, Bartosz,
  Bastien, Bates, Baumgarten, Bautista, Beaton, Beers, Belfiore, Bender,
  Bernardi, Bershady, Beutler, Bird, Bizyaev, Blanc, Blanton, Blomqvist,
  Bolton, Boquien, Borissova, Bovy, Andres Bradna~Diaz, Nielsen~Brandt,
  Brinkmann, Brownstein, Bundy, Burgasser, Burtin, Busca, Ca{\~n}as,
  Cano-D{\'{\i}}az, Cappellari, Carrera, Casey, Cervantes~Sodi, Chen, Cherinka,
  Chiappini, Doohyun~Choi, Chojnowski, Chuang, Chung, Clerc, Cohen, Comerford,
  Comparat, Correa~do Nascimento, da~Costa, Cousinou, Covey, Crane,
  Cruz-Gonzalez, Cunha, da~Silva~Ilha, Damke, Darling, Davidson, Dawson,
  de~Icaza~Lizaola, de~la Macorra, de~la Torre, De~Lee, de~Sainte~Agathe,
  Deconto~Machado, Dell{'}Agli, Delubac, Diamond-Stanic, Donor, Jos{\'e
  Downes}, Drory, du~Mas~des Bourboux, Duckworth, Dwelly, Dyer, Ebelke,
  Davis~Eigenbrot, Eisenstein, Elsworth, Emsellem, Eracleous, Erfanianfar,
  Escoffier, Fan, Fern{\'a}ndez~Alvar, Fernandez-Trincado, Cirolini, Feuillet,
  Finoguenov, Fleming, Font-Ribera, Freischlad, Frinchaboy, Fu, G{\'o}mez
  Maqueo~Chew, Galbany, Garc{\'{\i}}a~P{\'e}rez, Garcia-Dias,
  Garc{\'{\i}}a-Hern{\'a}ndez, Garma~Oehmichen, Gaulme, Gelfand,
  Gil-Mar{\'{\i}}n, Gillespie, Goddard, Gonz{\'a}lez~Hern{\'a}ndez,
  Gonzalez-Perez, Grabowski, Green, Grier, Gueguen, Guo, Guy, Hagen, Hall,
  Harding, Hasselquist, Hawley, Hayes, Hearty, Hekker, Hernandez,
  Hernandez~Toledo, Hogg, Holley-Bockelmann, Holtzman, Hou, Hsieh, Hunt,
  Hutchinson, Hwang, Jimenez~Angel, Johnson, Jones, J{\"o}nsson, Jullo,
  Sakil~Khan, Kinemuchi, Kirkby, Kirkpatrick, Kitaura, Knapp, Kneib, Kollmeier,
  Lacerna, Lane, Lang, Law, Le~Goff, Lee, Li, Li, Lian, Liang, Lima, Lin, Long,
  Lucatello, Lundgren, Mackereth, MacLeod, Mahadevan, Geimba~Maia, Majewski,
  Manchado, Maraston, Mariappan, Marques-Chaves, Masseron, Masters, McDermid,
  McGreer, Melendez, Meneses-Goytia, Merloni, Merrifield, Meszaros, Meza,
  Minchev, Minniti, Mueller, Muller-Sanchez, Muna, Mu{\~n}oz, Myers, Nair,
  Nandra, Ness, Newman, Nichol, Nidever, Nitschelm, Noterdaeme, O{'}Connell,
  Oelkers, Oravetz, Oravetz, Aquino~Ort{\'{\i}}z, Osorio, Pace, Padilla,
  Palanque-Delabrouille, Alonso~Palicio, Pan, Pan, Parikh, P{\^a}ris, Park,
  Peirani, Pellejero-Ibanez, Penny, Percival, Perez-Fournon, Petitjean, Pieri,
  Pinsonneault, Pisani, Prada, Prakash, Queiroz, Raddick, Raichoor,
  Barboza~Rembold, Richstein, Riffel, Riffel, Rix, Robin,
  Rodr{\'{\i}}guez~Torres, Rom{\'a}n-Z{\'u}{\~n}iga, Ross, Rossi, Ruan,
  Ruggeri, Ruiz, Salvato, S{\'a}nchez, S{\'a}nchez, Sanchez~Almeida,
  S{\'a}nchez-Gallego, Santana~Rojas, Santiago, Schiavon, Schimoia, Schlafly,
  Schlegel, Schneider, Schuster, Schwope, Seo, Serenelli, Shen, Shen, Shetrone,
  Shull, Silva~Aguirre, Simon, Skrutskie, Slosar, Smethurst, Smith, Sobeck,
  Somers, Souter, Souto, Spindler, Stark, Stassun, Steinmetz, Stello,
  Storchi-Bergmann, Streblyanska, Stringfellow, Su{\'a}rez, Sun, Szigeti,
  Taghizadeh-Popp, Talbot, Tang, Tao, Tayar, Tembe, Teske, Thakar, Thomas,
  Tissera, Tojeiro, Tremonti, Troup, Urry, Valenzuela, van~den Bosch,
  Vargas-Gonz{\'a}lez, Vargas-Maga{\~n}a, Vazquez, Villanova, Vogt, Wake, Wang,
  Weaver, Weijmans, Weinberg, Westfall, Whelan, Wilcots, Wild, Williams,
  Wilson, Wood-Vasey, Wylezalek, Xiao, Yan, Yang, Ybarra, Y{\`e}che, Zakamska,
  Zamora, Zarrouk, Zasowski, Zhang, Zhao, Zhao, Zheng, Zheng, Zhou, Zhu, Zinn,
  \& Zou}]{Abolfathi2018}
Abolfathi, B., Aguado, D.~S., Aguilar, G., {et~al.} 2018,
  \href{http://dx.doi.org/10.3847/1538-4365/aa9e8a}{\JournalTitle{\apjs}, 235,
  42}

\bibitem[{Baldwin {et~al.}(1981)Baldwin, Phillips, \& Terlevich}]{Baldwin1981}
Baldwin, J.~A., Phillips, M.~M., \& Terlevich, R. 1981,
  \href{http://dx.doi.org/10.1086/130766}{\JournalTitle{\pasp}, 93, 5}

\bibitem[{Balogh {et~al.}(1999)Balogh, Morris, Yee, Carlberg, \&
  Ellingson}]{Balogh1999}
Balogh, M.~L., Morris, S.~L., Yee, H. K.~C., Carlberg, R.~G., \& Ellingson, E.
  1999, \href{http://dx.doi.org/10.1086/308056}{\JournalTitle{\apj}, 527, 54}

\bibitem[{{Battisti} {et~al.}(2016){Battisti}, {Calzetti}, \&
  {Chary}}]{Battisti2016}
{Battisti}, A.~J., {Calzetti}, D., \& {Chary}, R.-R. 2016,
  \href{http://dx.doi.org/10.3847/0004-637X/818/1/13}{\JournalTitle{\apj}, 818,
  13}

\bibitem[{Battisti {et~al.}(2017{\natexlab{a}})Battisti, Calzetti, \&
  Chary}]{Battisti2017a}
Battisti, A.~J., Calzetti, D., \& Chary, R.-R. 2017{\natexlab{a}},
  \href{http://dx.doi.org/10.3847/1538-4357/aa9a43}{\JournalTitle{\apj}, 851,
  90}

\bibitem[{Battisti {et~al.}(2017{\natexlab{b}})Battisti, Calzetti, \&
  Chary}]{Battisti2017}
---. 2017{\natexlab{b}},
  \href{http://dx.doi.org/10.3847/1538-4357/aa6fb2}{\JournalTitle{\apj}, 840,
  109}

\bibitem[{Blanton {et~al.}(2011)Blanton, Kazin, Muna, Weaver, \&
  Price-Whelan}]{Blanton2011}
Blanton, M.~R., Kazin, E., Muna, D., Weaver, B.~A., \& Price-Whelan, A. 2011,
  \href{http://dx.doi.org/10.1088/0004-6256/142/1/31}{\JournalTitle{\aj}, 142,
  31}

\bibitem[{{Bruzual} \& {Charlot}(2003)}]{Bruzual2003}
{Bruzual}, G., \& {Charlot}, S. 2003,
  \href{http://dx.doi.org/10.1046/j.1365-8711.2003.06897.x}{\JournalTitle{\mnras},
  344, 1000}

\bibitem[{Bundy {et~al.}(2015)Bundy, Bershady, Law, Yan, Drory, MacDonald,
  Wake, Cherinka, S{\'a}nchez-Gallego, Weijmans, Thomas, Tremonti, Masters,
  Coccato, Diamond-Stanic, Arag{\'o}n-Salamanca, Avila-Reese, Badenes,
  Falc{\'o}n-Barroso, Belfiore, Bizyaev, Blanc, Bland-Hawthorn, Blanton,
  Brownstein, Byler, Cappellari, Conroy, Dutton, Emsellem, Etherington,
  Frinchaboy, Fu, Gunn, Harding, Johnston, Kauffmann, Kinemuchi, Klaene,
  Knapen, Leauthaud, Li, Lin, Maiolino, Malanushenko, Malanushenko, Mao,
  Maraston, McDermid, Merrifield, Nichol, Oravetz, Pan, Parejko, Sanchez,
  Schlegel, Simmons, Steele, Steinmetz, Thanjavur, Thompson, Tinker, van~den
  Bosch, Westfall, Wilkinson, Wright, Xiao, \& Zhang}]{Bundy2015}
Bundy, K., Bershady, M.~A., Law, D.~R., {et~al.} 2015,
  \href{http://dx.doi.org/10.1088/0004-637X/798/1/7}{\JournalTitle{\apj}, 798,
  7}

\bibitem[{Calzetti(1997)}]{Calzetti1997}
Calzetti, D. 1997, \href{http://dx.doi.org/10.1063/1.53764}{in American
  Institute of Physics Conference Series, Vol. 408, The Ultraviolet Universe at
  Low and High Redshift, ed. W.~H. {Waller}}, 403

\bibitem[{Calzetti(2001)}]{Calzetti2001}
Calzetti, D. 2001,
  \href{http://dx.doi.org/10.1086/324269}{\JournalTitle{\pasp}, 113, 1449}

\bibitem[{{Calzetti} {et~al.}(2000){Calzetti}, {Armus}, {Bohlin}, {Kinney},
  {Koornneef}, \& {Storchi-Bergmann}}]{Calzetti2000}
{Calzetti}, D., {Armus}, L., {Bohlin}, R.~C., {et~al.} 2000,
  \href{http://dx.doi.org/10.1086/308692}{\JournalTitle{\apj}, 533, 682}

\bibitem[{{Calzetti} {et~al.}(1994){Calzetti}, {Kinney}, \&
  {Storchi-Bergmann}}]{Calzetti1994}
{Calzetti}, D., {Kinney}, A.~L., \& {Storchi-Bergmann}, T. 1994,
  \href{http://dx.doi.org/10.1086/174346}{\JournalTitle{\apj}, 429, 582}

\bibitem[{{Cardelli} {et~al.}(1989){Cardelli}, {Clayton}, \&
  {Mathis}}]{Cardelli1989}
{Cardelli}, J.~A., {Clayton}, G.~C., \& {Mathis}, J.~S. 1989,
  \href{http://dx.doi.org/10.1086/167900}{\JournalTitle{\apj}, 345, 245}

\bibitem[{Chabrier(2003)}]{Chabrier2003}
Chabrier, G. 2003,
  \href{http://dx.doi.org/10.1086/376392}{\JournalTitle{\pasp}, 115, 763}

\bibitem[{{Charlot} \& {Fall}(2000)}]{Charlot2000}
{Charlot}, S., \& {Fall}, S.~M. 2000,
  \href{http://dx.doi.org/10.1086/309250}{\JournalTitle{\apj}, 539, 718}

\bibitem[{Cid~Fernandes {et~al.}(2005)Cid~Fernandes, Mateus, Sodr{\'e},
  Stasi{\'n}ska, \& Gomes}]{CidFernandes2005}
Cid~Fernandes, R., Mateus, A., Sodr{\'e}, L., Stasi{\'n}ska, G., \& Gomes,
  J.~M. 2005,
  \href{http://dx.doi.org/10.1111/j.1365-2966.2005.08752.x}{\JournalTitle{\mnras},
  358, 363}

\bibitem[{Clayton {et~al.}(2015)Clayton, Gordon, Bianchi, Massa, Fitzpatrick,
  Bohlin, \& Wolff}]{Clayton2015}
Clayton, G.~C., Gordon, K.~D., Bianchi, L.~C., {et~al.} 2015,
  \href{http://dx.doi.org/10.1088/0004-637X/815/1/14}{\JournalTitle{\apj}, 815,
  14}

\bibitem[{Conroy(2013)}]{Conroy2013}
Conroy, C. 2013,
  \href{http://dx.doi.org/10.1146/annurev-astro-082812-141017}{\JournalTitle{\araa},
  51, 393}

\bibitem[{Cullen {et~al.}(2018)Cullen, McLure, Khochfar, Dunlop, Dalla~Vecchia,
  Carnall, Bourne, Castellano, Cimatti, Cirasuolo, Elbaz, Fynbo, Garilli,
  Koekemoer, Marchi, Pentericci, Talia, \& Zamorani}]{Cullen2018}
Cullen, F., McLure, R.~J., Khochfar, S., {et~al.} 2018,
  \href{http://dx.doi.org/10.1093/mnras/sty469}{\JournalTitle{\mnras}, 476,
  3218}

\bibitem[{Draine(2003)}]{Draine2003}
Draine, B.~T. 2003,
  \href{http://dx.doi.org/10.1146/annurev.astro.41.011802.094840}{\JournalTitle{\araa},
  41, 241}

\bibitem[{Flores-Fajardo {et~al.}(2011)Flores-Fajardo, Morisset, Stasi{\'n}ska,
  \& Binette}]{Flores-Fajardo2011}
Flores-Fajardo, N., Morisset, C., Stasi{\'n}ska, G., \& Binette, L. 2011,
  \href{http://dx.doi.org/10.1111/j.1365-2966.2011.18848.x}{\JournalTitle{\mnras},
  415, 2182}

\bibitem[{Gordon {et~al.}(2003)Gordon, Clayton, Misselt, Landolt, \&
  Wolff}]{Gordon2003}
Gordon, K.~D., Clayton, G.~C., Misselt, K.~A., Landolt, A.~U., \& Wolff, M.~J.
  2003, \href{http://dx.doi.org/10.1086/376774}{\JournalTitle{\apj}, 594, 279}

\bibitem[{Gusev(2014)}]{Gusev2014}
Gusev, A.~S. 2014,
  \href{http://dx.doi.org/10.1093/mnras/stu1095}{\JournalTitle{\mnras}, 442,
  3711}

\bibitem[{Haffner {et~al.}(2009)Haffner, Dettmar, Beckman, Wood, Slavin,
  Giammanco, Madsen, Zurita, \& Reynolds}]{Haffner2009}
Haffner, L.~M., Dettmar, R.-J., Beckman, J.~E., {et~al.} 2009,
  \href{http://dx.doi.org/10.1103/RevModPhys.81.969}{\JournalTitle{Reviews of
  Modern Physics}, 81, 969}

\bibitem[{Hsieh {et~al.}(2017)Hsieh, Lin, Lin, Pan, Hsu, S{\'a}nchez,
  Cano-D{\'{\i}}az, Zhang, Yan, Barrera-Ballesteros, Boquien, Riffel,
  Brownstein, Cruz-Gonz{\'a}lez, Hagen, Ibarra, Pan, Bizyaev, Oravetz, \&
  Simmons}]{Hsieh2017}
Hsieh, B.~C., Lin, L., Lin, J.~H., {et~al.} 2017,
  \href{http://dx.doi.org/10.3847/2041-8213/aa9d80}{\JournalTitle{\apjl}, 851,
  L24}

\bibitem[{{Kauffmann} {et~al.}(2003a){Kauffmann}, {Heckman}, {Tremonti},
  {Brinchmann}, {Charlot}, {White}, {Ridgway}, {Brinkmann}, {Fukugita}, {Hall},
  {Ivezi{\'c}}, {Richards}, \& {Schneider}}]{Kauffmann2003a}
{Kauffmann}, G., {Heckman}, T.~M., {Tremonti}, C., {et~al.} 2003a,
  \href{http://dx.doi.org/10.1111/j.1365-2966.2003.07154.x}{\JournalTitle{\mnras},
  346, 1055}

\bibitem[{Kauffmann {et~al.}(2003b)Kauffmann, Heckman, White, Charlot,
  Tremonti, Brinchmann, Bruzual, Peng, Seibert, Bernardi, Blanton, Brinkmann,
  Castander, Cs{\'a}bai, Fukugita, Ivezic, Munn, Nichol, Padmanabhan, Thakar,
  Weinberg, \& York}]{Kauffmann2003}
Kauffmann, G., Heckman, T.~M., White, S. D.~M., {et~al.} 2003b,
  \href{http://dx.doi.org/10.1046/j.1365-8711.2003.06291.x}{\JournalTitle{\mnras},
  341, 33}

\bibitem[{Kewley {et~al.}(2001)Kewley, Dopita, Sutherland, Heisler, \&
  Trevena}]{Kewley2001}
Kewley, L.~J., Dopita, M.~A., Sutherland, R.~S., Heisler, C.~A., \& Trevena, J.
  2001, \href{http://dx.doi.org/10.1086/321545}{\JournalTitle{\apj}, 556, 121}

\bibitem[{Kewley \& Ellison(2008)}]{Kewley2008}
Kewley, L.~J., \& Ellison, S.~L. 2008,
  \href{http://dx.doi.org/10.1086/587500}{\JournalTitle{\apj}, 681, 1183}

\bibitem[{Kewley {et~al.}(2006)Kewley, Groves, Kauffmann, \&
  Heckman}]{Kewley2006}
Kewley, L.~J., Groves, B., Kauffmann, G., \& Heckman, T. 2006,
  \href{http://dx.doi.org/10.1111/j.1365-2966.2006.10859.x}{\JournalTitle{\mnras},
  372, 961}

\bibitem[{Kinney {et~al.}(1994)Kinney, Calzetti, Bica, \&
  Storchi-Bergmann}]{Kinney1994}
Kinney, A.~L., Calzetti, D., Bica, E., \& Storchi-Bergmann, T. 1994,
  \href{http://dx.doi.org/10.1086/174309}{\JournalTitle{\apj}, 429, 172}

\bibitem[{Kreckel {et~al.}(2013)Kreckel, Groves, Schinnerer, Johnson, Aniano,
  Calzetti, Croxall, Draine, Gordon, Crocker, Dale, Hunt, Kennicutt, Meidt,
  Smith, \& Tabatabaei}]{Kreckel2013}
Kreckel, K., Groves, B., Schinnerer, E., {et~al.} 2013,
  \href{http://dx.doi.org/10.1088/0004-637X/771/1/62}{\JournalTitle{\apj}, 771,
  62}

\bibitem[{Lacerda {et~al.}(2018)Lacerda, Cid~Fernandes, Couto, Stasi{\'n}ska,
  Garc{\'{\i}}a-Benito, Vale~Asari, P{\'e}rez, Gonz{\'a}lez~Delgado,
  S{\'a}nchez, \& de~Amorim}]{Lacerda2018}
Lacerda, E. A.~D., Cid~Fernandes, R., Couto, G.~S., {et~al.} 2018,
  \href{http://dx.doi.org/10.1093/mnras/stx3022}{\JournalTitle{\mnras}, 474,
  3727}

\bibitem[{Law {et~al.}(2016)Law, Cherinka, Yan, Andrews, Bershady, Bizyaev,
  Blanc, Blanton, Bolton, Brownstein, Bundy, Chen, Drory, D'Souza, Fu, Jones,
  Kauffmann, MacDonald, Masters, Newman, Parejko, S{\'a}nchez-Gallego,
  S{\'a}nchez, Schlegel, Thomas, Wake, Weijmans, Westfall, \& Zhang}]{Law2016}
Law, D.~R., Cherinka, B., Yan, R., {et~al.} 2016,
  \href{http://dx.doi.org/10.3847/0004-6256/152/4/83}{\JournalTitle{\aj}, 152,
  83}

\bibitem[{Li {et~al.}(2015)Li, Wang, Lin, Bershady, Bundy, Tremonti, Xiao, Yan,
  Bizyaev, Blanton, Cales, Cherinka, Cheung, Drory, Emsellem, Fu, Gelfand, Law,
  Lin, MacDonald, Maraston, Masters, Merrifield, Pan, S{\'a}nchez, Schneider,
  Thomas, Wake, Wang, Weijmans, Wilkinson, Yoachim, Zhang, \& Zheng}]{Li2015}
Li, C., Wang, E., Lin, L., {et~al.} 2015,
  \href{http://dx.doi.org/10.1088/0004-637X/804/2/125}{\JournalTitle{\apj},
  804, 125}

\bibitem[{Lin \& Kong(2020)}]{Lin2020}
Lin, Z., \& Kong, X. 2020,
  \href{http://dx.doi.org/10.3847/1538-4357/ab5f0e}{\JournalTitle{\apj}, 888,
  88}

\bibitem[{Lin {et~al.}(2017)Lin, Hu, Kong, Gao, Zou, Wang, Cheng, Fang, Lin, \&
  Wang}]{LinZ2017}
Lin, Z., Hu, N., Kong, X., {et~al.} 2017,
  \href{http://dx.doi.org/10.3847/1538-4357/aa6f14}{\JournalTitle{\apj}, 842,
  97}

\bibitem[{Liu {et~al.}(2013)Liu, Calzetti, Hong, Whitmore, Chandar, O'Connell,
  Blair, Cohen, Frogel, \& Kim}]{Liu2013b}
Liu, G., Calzetti, D., Hong, S., {et~al.} 2013,
  \href{http://dx.doi.org/10.1088/2041-8205/778/2/L41}{\JournalTitle{\apjl},
  778, L41}

\bibitem[{Liu {et~al.}(2018)Liu, Wang, Lin, Gao, Liu, Berhane~Teklu, \&
  Kong}]{Liu2018}
Liu, Q., Wang, E., Lin, Z., {et~al.} 2018,
  \href{http://dx.doi.org/10.3847/1538-4357/aab3d5}{\JournalTitle{\apj}, 857,
  17}

\bibitem[{Lo~Faro {et~al.}(2017)Lo~Faro, Buat, Roehlly, Alvarez-Marquez,
  Burgarella, Silva, \& Efstathiou}]{LoFaro2017}
Lo~Faro, B., Buat, V., Roehlly, Y., {et~al.} 2017,
  \href{http://dx.doi.org/10.1093/mnras/stx1901}{\JournalTitle{\mnras}, 472,
  1372}

\bibitem[{Ly {et~al.}(2014)Ly, Malkan, Nagao, Kashikawa, Shimasaku, \&
  Hayashi}]{Ly2014}
Ly, C., Malkan, M.~A., Nagao, T., {et~al.} 2014,
  \href{http://dx.doi.org/10.1088/0004-637X/780/2/122}{\JournalTitle{\apj},
  780, 122}

\bibitem[{Marino {et~al.}(2013)Marino, Rosales-Ortega, S{\'a}nchez, Gil~de Paz,
  V{\'{\i}}lchez, Miralles-Caballero, Kehrig, P{\'e}rez-Montero, Stanishev,
  Iglesias-P{\'a}ramo, D{\'{\i}}az, Castillo-Morales, Kennicutt,
  L{\'o}pez-S{\'a}nchez, Galbany, Garc{\'{\i}}a-Benito, Mast, Mendez-Abreu,
  Monreal-Ibero, Husemann, Walcher, Garc{\'{\i}}a-Lorenzo, Masegosa, Del
  Olmo~Orozco, Mour{\~a}o, Ziegler, Moll{\'a}, Papaderos,
  S{\'a}nchez-Bl{\'a}zquez, Gonz{\'a}lez~Delgado, Falc{\'o}n-Barroso, Roth,
  van~de Ven, \& Team}]{Marino2013}
Marino, R.~A., Rosales-Ortega, F.~F., S{\'a}nchez, S.~F., {et~al.} 2013,
  \href{http://dx.doi.org/10.1051/0004-6361/201321956}{\JournalTitle{\aap},
  559, A114}

\bibitem[{Markwardt(2009)}]{Markwardt2009}
Markwardt, C.~B. 2009,
  \href{http://adsabs.harvard.edu/abs/2009ASPC..411..251M}{in Astronomical
  Society of the Pacific Conference Series, Vol. 411, Astronomical Data
  Analysis Software and Systems XVIII, ed. D.~A. {Bohlender}, D.~{Durand}, \&
  P.~{Dowler}}, 251

\bibitem[{Misselt {et~al.}(1999)Misselt, Clayton, \& Gordon}]{Misselt1999}
Misselt, K.~A., Clayton, G.~C., \& Gordon, K.~D. 1999,
  \href{http://dx.doi.org/10.1086/307010}{\JournalTitle{\apj}, 515, 128}

\bibitem[{Narayanan {et~al.}(2018)Narayanan, Conroy, Dav{\'e}, Johnson, \&
  Popping}]{Narayanan2018}
Narayanan, D., Conroy, C., Dav{\'e}, R., Johnson, B.~D., \& Popping, G. 2018,
  \href{http://dx.doi.org/10.3847/1538-4357/aaed25}{\JournalTitle{\apj}, 869,
  70}

\bibitem[{{Reddy} {et~al.}(2015){Reddy}, {Kriek}, {Shapley}, {Freeman},
  {Siana}, {Coil}, {Mobasher}, {Price}, {Sanders}, \& {Shivaei}}]{Reddy2015}
{Reddy}, N.~A., {Kriek}, M., {Shapley}, A.~E., {et~al.} 2015,
  \href{http://dx.doi.org/10.1088/0004-637X/806/2/259}{\JournalTitle{\apj},
  806, 259}

\bibitem[{Salim {et~al.}(2018)Salim, Boquien, \& Lee}]{Salim2018}
Salim, S., Boquien, M., \& Lee, J.~C. 2018,
  \href{http://dx.doi.org/10.3847/1538-4357/aabf3c}{\JournalTitle{\apj}, 859,
  11}

\bibitem[{Salmon {et~al.}(2016)Salmon, Papovich, Long, Willner, Finkelstein,
  Ferguson, Dickinson, Duncan, Faber, Hathi, Koekemoer, Kurczynski, Newman,
  Pacifici, P{\'e}rez-Gonz{\'a}lez, \& Pforr}]{Salmon2016}
Salmon, B., Papovich, C., Long, J., {et~al.} 2016,
  \href{http://dx.doi.org/10.3847/0004-637X/827/1/20}{\JournalTitle{\apj}, 827,
  20}

\bibitem[{{Scoville} {et~al.}(2015){Scoville}, {Faisst}, {Capak}, {Kakazu},
  {Li}, \& {Steinhardt}}]{Scoville2015}
{Scoville}, N., {Faisst}, A., {Capak}, P., {et~al.} 2015,
  \href{http://dx.doi.org/10.1088/0004-637X/800/2/108}{\JournalTitle{\apj},
  800, 108}

\bibitem[{Storey \& Hummer(1995)}]{Storey1995}
Storey, P.~J., \& Hummer, D.~G. 1995,
  \href{http://dx.doi.org/10.1093/mnras/272.1.41}{\JournalTitle{\mnras}, 272,
  41}

\bibitem[{Tress {et~al.}(2018)Tress, M{\'a}rmol-Queralt{\'o}, Ferreras,
  P{\'e}rez-Gonz{\'a}lez, Barro, Pampliega, Cava, Dom{\'{\i}}nguez-S{\'a}nchez,
  Eliche-Moral, Espino-Briones, Esquej, Hern{\'a}n-Caballero, Rodighiero, \&
  Rodriguez-Mu{\~n}oz}]{Tress2018}
Tress, M., M{\'a}rmol-Queralt{\'o}, E., Ferreras, I., {et~al.} 2018,
  \href{http://dx.doi.org/10.1093/mnras/stx3334}{\JournalTitle{\mnras}, 475,
  2363}

\bibitem[{Viaene {et~al.}(2017)Viaene, Sarzi, Baes, Fritz, \&
  Puerari}]{Viaene2017}
Viaene, S., Sarzi, M., Baes, M., Fritz, J., \& Puerari, I. 2017,
  \href{http://dx.doi.org/10.1093/mnras/stx1781}{\JournalTitle{\mnras}, 472,
  1286}

\bibitem[{Viaene {et~al.}(2019)Viaene, Sarzi, Zabel, Coccato, Corsini, Davis,
  De~Vis, de~Zeeuw, Falc{\'o}n-Barroso, Gadotti, Iodice, Lyubenova, McDermid,
  Morelli, Nedelchev, Pinna, Spriggs, \& van~de Ven}]{Viaene2019}
Viaene, S., Sarzi, M., Zabel, N., {et~al.} 2019,
  \href{http://dx.doi.org/10.1051/0004-6361/201834465}{\JournalTitle{\aap},
  622, A89}

\bibitem[{Weingartner \& Draine(2001)}]{Weingartner2001}
Weingartner, J.~C., \& Draine, B.~T. 2001,
  \href{http://dx.doi.org/10.1086/318651}{\JournalTitle{\apj}, 548, 296}

\bibitem[{Wild {et~al.}(2011)Wild, Charlot, Brinchmann, Heckman, Vince,
  Pacifici, \& Chevallard}]{Wild2011}
Wild, V., Charlot, S., Brinchmann, J., {et~al.} 2011,
  \href{http://dx.doi.org/10.1111/j.1365-2966.2011.19367.x}{\JournalTitle{\mnras},
  417, 1760}

\bibitem[{Yip {et~al.}(2010)Yip, Szalay, Wyse, Dobos, Budav{\'a}ri, \&
  Csabai}]{Yip2010}
Yip, C.-W., Szalay, A.~S., Wyse, R. F.~G., {et~al.} 2010,
  \href{http://dx.doi.org/10.1088/0004-637X/709/2/780}{\JournalTitle{\apj},
  709, 780}

\bibitem[{Zhang {et~al.}(2017)Zhang, Yan, Bundy, Bershady, Haffner, Walterbos,
  Maiolino, Tremonti, Thomas, Drory, Jones, Belfiore, S{\'a}nchez,
  Diamond-Stanic, Bizyaev, Nitschelm, Andrews, Brinkmann, Brownstein, Cheung,
  Li, Law, Roman~Lopes, Oravetz, Pan, Storchi~Bergmann, \& Simmons}]{Zhang2017}
Zhang, K., Yan, R., Bundy, K., {et~al.} 2017,
  \href{http://dx.doi.org/10.1093/mnras/stw3308}{\JournalTitle{\mnras}, 466,
  3217}

\end{thebibliography}

\end{CJK*}
\end{document}